\documentclass[12pt]{iopart}

\usepackage{graphicx}
\usepackage{amssymb}
\usepackage{lscape}
\usepackage{rotating}

\begin{document}

\title[]{Nodal Structure of Unconventional Superconductors Probed by the
Angle Resolved Thermal Transport Measurements}

\author{Y Matsuda$^{1,2}$, K Izawa$^{2,3}$, and I Vekhter$^{4}$}

\address{$^1$ Department of Physics, Kyoto University, Kyoto 606-8502, Japan}
\address{$^2$ Institute for Solid State Physics, University of Tokyo, Kashiwanoha, Kashiwa, Chiba 277-8581, Japan}
\address{$^3$ DRFMC/SPSMS/LCP, CEA-Grenoble, 17 rue des Martyrs 38054, Grenoble cedex9 France}
\address{$^4$ Department of Physics and Astronomy, Louisiana State University, Baton Rouge, LA 70803, USA}

\ead{matsuda@scphys.kyoto-u.ac.jp}
\begin{abstract}
Over the past two decades, unconventional superconductivity with
gap symmetry other than $s$-wave has been found in several classes
of materials, including heavy fermion (HF), high-$T_c$, and
organic superconductors. Unconventional superconductivity is
characterized by anisotropic superconducting gap functions, which
may have zeros (nodes) along certain directions in the Brillouin
zone. The nodal structure is closely related to the pairing
interaction, and it is widely believed that the presence of nodes
is a signature of magnetic or some other exotic, rather than
conventional phonon-mediated, pairing mechanism. Therefore
experimental determination of the gap function is of fundamental
importance. However,  the detailed gap structure, especially the
direction of the nodes, is an unresolved issue in most
unconventional superconductors. Recently it has been demonstrated
that the thermal conductivity and specific heat measurements under
magnetic field rotated relative to the crystal axes are a powerful
method for determining the shape of the gap and the nodal
directions in the bulk. Here we review the theoretical
underpinnings of the method and the results for the nodal
structure of several unconventional superconductors, including
borocarbide YNi$_2$B$_2$C, heavy fermions UPd$_2$Al$_3$,
CeCoIn$_5$, and PrOs$_4$Sb$_{12}$, organic superconductor,
$\kappa$-(BEDT-TTF)$_2$Cu(NCS)$_2$, and ruthenate Sr$_2$RuO$_4$,
determined by angular variation of the thermal conductivity and
heat capacity.

\end{abstract}

\maketitle

\section{Introduction}

Superconducting transition is a second order phase transition
associated with a spontaneous symmetry breaking. Consequently the
order parameter that appears below the transition temperature,
$T_c$, characterizes the lowering of the symmetry in the low
temperature ordered phase compared to the metallic state. The
order parameter is also related to the gap in the single particle
excitation spectrum, and hence its symmetry is reflected in the
elementary excitations in the superconducting phase. These, in
turn, determine the observed transport and thermodynamic
properties \cite{sigrist,mineev}.

At the microscopic level, the symmetry of the order parameter is
intimately related to the pairing interaction. Therefore the
identification of this symmetry is of central importance in the
study of the superconductivity, especially in novel correlated
electron materials. This review describes recent progress and
current status of the efforts to use transport properties as a
reliable tool for the determination of the gap symmetry in
unconventional superconductors in the bulk.

In all superconductors the gauge symmetry is broken below the
transition temperature. At the same time, in most materials, the
energy gap has the full spatial symmetry of the underlying crystal
lattice. In the simplest cases, this corresponds to a gap
isotropic in the momentum space, i.e. independent of the
directions at the Fermi surface. These superconductors are termed
conventional, or $s$-wave. However, it is also possible that the
spatial symmetry of the superconducting order parameter is {\em
lower} than that of the lattice, and such superconductors are
labeled unconventional. Sec.\ref{sec:UnconvGap} presents the
detailed symmetry classification of the unconventional
superconductors.  These materials first appeared in 1979, when the
heavy fermion CeCu$_2$Si$_2$ was discovered \cite{steglich}, and
by now superconductivity with non $s$-wave symmetry is ubiquitous.
Examples of it include anisotropic gaps with zeroes (nodes), odd
parity superconducting condensate wave functions, and broken time
reversal symmetry. Realization of these possibilities is proved or
strongly suggested in several classes of materials, including
heavy fermion \cite{stewart1984,volovik1985,thalmeier}, high-$T_c$
cuprates \cite{tsuei}, ruthenate\cite{maeno1994,bergemann,maeno},
cobaltate \cite{takada}, intermetallic compounds\cite{canfield},
and organic
superconductors\cite{jerome1980,kanoda,ross,kuroki2006}.

In these systems, strong electron correlations often give rise to
the Cooper pair states with a non-zero angular momentum.
Unfortunately, the experimental determination of the detailed
superconducting gap structure is an extremely difficult task. The
phase sensitive measurements testing the sign change of the gap
have only been done in the high-$T_c$ cuprates, firmly
establishing (along with a number of other techniques probing the
gap anisotropy) the predominant $d_{x^2-y^2}$ pairing symmetry
\cite{tsuei}.

Unconventional superconductivity often occurs in heavy fermion
compounds, containing $f$-electrons (lanthanide (4f) and actinide
(5f)), especially in materials containing Ce, Pr, U and Pu atoms.
At high temperature $f$-electrons are essentially localized with
well-defined magnetic moments. As the temperature is lowered, the
$f$-electrons begin to delocalize due to the hybridization with
conduction electron band($s,p,d$-orbital), and Kondo screening. At
yet lower $T$, the $f$-electrons become itinerant, forming a
narrow conduction band with heavy effective electron mass (up to a
few hundred to a thousand times the free electron mass).  Strong
Coulomb repulsion within a narrow band and the magnetic
interaction between remnant unscreened 4$f$ or 5$f$ moments leads
to notable many-body effects, and, likely, to superconductivity
mediated by magnetic fluctuations.  These effects are believed to
be especially pronounced in the vicinity of zero-temperature
magnetic instability (quantum critical point (QCP)).    Although
the superconducting gap is thought to be unconventional in most
heavy fermion superconductors, its detailed momentum dependence is
an unresolved issue \cite{stewart1984,volovik1985,thalmeier}.

Since the discovery of superconductivity in  organic materials
about two decades ago, superconductivity has been reported in more
than 100 organic compounds.  Among them, two families of
compounds, quasi-1D Bechgaard salts (TMTSF)$_2$X (X=ClO$_4$,
PF$_6$, AsF$_6$, etc.) and quasi-2D $\kappa$-(BEDT-TTF)$_2$X salts
($\kappa$-(ET)$_2$X), where the ion X can, for example, be
Cu(SCN)$_2$, Cu[N(CN)$_2$]Br or I$_3$, are particularly prominent
candidates for unconventional pairing.  The pairing symmetry in
both systems is still undetermined, and is one of the most
intriguing problems in the field \cite{kanoda,ross,kuroki2006}.

The unconventional superconductivity has also been reported in
some of transition metal oxides other than high-$T_c$ cuprates.
Especially, the superconducting  gap functions in layered
ruthenate Sr$_2$RuO$_4$\cite{bergemann,maeno} and  layered
cobaltate Na$_x$CoO$_2 \cdot $yH$_2$O\cite{takada} have attracted
considerable interest.    Moreover, among  intermetallic compound,
the gap function of borocarbide superconductors (Y,
Lu)Ni$_2$B$_2$C \cite{canfield} has been reported to be very
anisotropic, which implies that the simple electron-phonon pairing
mechanism originally envisaged for these compounds is not the sole
source for pairing interaction.

Evidence for anisotropic gap in a variety of systems has continued
to motivate theorists to propose new models for the unconventional
superconductivity, which make specific predictions for the shape
of the superconducting gap in momentum space. Experimental
determination of the gap symmetry is therefore of crucial
importance. The transport measurements are not, per s\'e, phase
sensitive, and therefore cannot unequivocally determine the sign
change in the gap function. They are, however, an extremely
sensitive probe of the anisotropy of the gap amplitude in the
momentum space, and have been extensively used in the last few
years to determine the shape of the gap in many materials at the
forefront of modern research. Below we give an overview of these
efforts.

\section{unconventional pairing state}
\label{sec:UnconvGap}

The general classification scheme for the superconducting order
parameter is based on its behavior under symmetry transformations.
The full symmetry group $\mathcal{G}$ of the crystal contains the
gauge group $U(1)$, crystal point group $G$, spin rotation group
$SU(2)$, and time reversal symmetry group $\mathcal{T}$,
\begin{equation}
\mathcal{G}=U(1) \otimes G \otimes SU(2) \otimes \mathcal{T}\, .
\end{equation}
The superconducting order breaks the $U(1)$ gauge symmetry below
$T_c$, and the simplest superconductors are those in which only
the $U(1)$ symmetry is broken; these are labeled conventional.
Unconventional superconductors break an additional symmetry
besides $U(1)$, and may include order parameters which
\begin{enumerate}
\item have odd  parity;
\item break time reversal symmetry;
\item break the point group symmetry of the crystal.
\end{enumerate}

The superconducting  order parameter is proportional to the gap
function $\Delta_{{\bf s}_1,{\bf s}_2}^{\ell}({\bf k})$, which, in
turn, is proportional to the amplitude of the wave function for a
Cooper pair $\Psi_{{\bf s}_1,{\bf s}_2}^{\ell}({\bf k})=\langle
\psi_{\bf k, {\bf s}_1} \psi_{-\bf k, {\bf s}_2}\rangle$. Here
${\bf k}$ is the quasiparticle momentum, ${\bf s}_i$ is the
electron spin, and $\psi$ is the electron annihilation operator.
The order parameter is called unconventional if it transforms
according to a nontrivial representation of the full symmetry
group. Pauli's exclusion principle requires $\Delta_{{\bf
s}_1,{\bf s}_2}^{\ell}({\bf k})$ to be antisymmetric under the the
particle interchange: $\Delta_{{\bf s}_1,{\bf s}_2}^{\ell}({\bf
k})=-\Delta_{{\bf s}_2,{\bf s}_1}^{\ell}({\bf -k})$. In the
simplest case of weak spin-orbit interaction, the total angular
momentum {\bf L} and total spin ${\bf S}={\bf s}_1+{\bf s}_2$ are
good quantum numbers, and   $\Delta_{{\bf s}_1,{\bf s}_2}({\bf
k})$ can be written as a product of orbital and spin parts,
\begin{equation}
\Delta_{{\bf s}_1,{\bf s}_2}^{\ell}({\bf k})=g_{\ell}({\bf k})
\chi_s ({\bf s}_1, {\bf s}_2).
\end{equation}
The orbital part, $g_{\ell}({\bf k})$, can be expanded in the
spherical harmonics $Y_{\ell m}(\widehat{\bf k})$, which are the
eigenfunctions of the angular momentum operator with the momentum
$\ell$ and its $z$-projections $m$,
\begin{equation}
g_{\ell}({\bf k})=\sum_{m=-\ell}^{\ell}a_{\ell m}(k) Y_{\ell m}(
\widehat{\bf k}).
\end{equation}
Here $\widehat{\bf k}={\bf k}/k_F$ denotes the direction of the
Fermi surface. $g_{\ell}({\bf k})$ is even for even values of
$\ell$ and odd for odd values of $\ell$, $g_{\ell}({\bf
k})=(-1)^{\ell}g_{\ell}({\bf -k})$, and superconductors with
$\ell=0,1,2,3,4,\ldots$ are labeled as having $s,p,d,f,g,\ldots$
wave gap respectively. This classification is valid for an
isotropic system. In a crystal, the spatial part of the Cooper
pair wave function is classified according to the irreducible
representations of the symmetry group of the lattice. However it
is common even in this case to refer to the possible pairing
states as having a particular angular momentum (rather than
belonging to a representation of the group with given symmetry
properties) and we use the notation here.

The spin part of the order parameter, $\chi_s({\bf s}_1,{\bf
s}_2)$, is a product of the spinors for the two electrons in the
Cooper pair. Therefore the gap function is a 2$\times$2 matrix in
spin space,
\begin{equation}
\bf\Delta^{\ell}_{\bf S}({\bf k})\equiv\bf\Delta^{\ell}_{{\bf
s}_1,{\bf s}_2}({\bf k})=\left(\begin{array}{cc}
\Delta_{\uparrow\uparrow}^{\ell}({\bf
k})&\Delta_{\uparrow\downarrow}^{\ell}({\bf
k})\\\Delta_{\downarrow\uparrow}^{\ell}({\bf
k})&\Delta_{\downarrow\downarrow}^{\ell}({\bf
k})\end{array}\right)
\end{equation}
In the singlet state, $S=0$, the spin part of the  wave function
is $|\uparrow\downarrow\rangle - |\downarrow\uparrow\rangle$, and
therefore the gap function is simply  proportional to the Pauli
matrix $\sigma_y$:
\begin{equation}
{\bf \Delta}_s({\bf k})=\Delta g_{\ell}({\bf k})i\sigma_y,
\end{equation}
where $\ell$ is even and $g_{\ell}$ is normalized.  The energy of
single particle excitations in this case is
\begin{equation}
E_{{\bf k}}=\sqrt{\xi_{\bf k}^2+\Delta^2|g_{\ell}({\bf k})|^2},
\label{eq:ExcitEnegy}
\end{equation}
where $\xi_{\bf k}$ is the band energy relative to the chemical
potential. For superconductors with an isotropic $\Delta({\bf k})$
the excitations have a finite energy gap everywhere at the Fermi
surface, while for anisotropic pairing the gap amplitude depends
on the components of $g({\bf k})$.

For spin triplet pairing $(S=1)$,  the wave function has
components corresponding to the three different spin projections,
$S^z$, on the quantization axis ($|{\uparrow\uparrow}\rangle,
|{\uparrow\downarrow}\rangle+|{\downarrow\uparrow}\rangle,
|{\downarrow\downarrow}\rangle)$. Consequently, it is common to
write the order parameter as
\begin{eqnarray}
{\bf \Delta}_t({\bf k})&=&i\Delta ({\bf d}({\bf k}){\bf \sigma})\sigma_y \nonumber \\
    &=& \Delta \left(\begin{array}{cc}-d_x({\bf k})+id_y({\bf k}) & d_z({\bf k})\\
  d_z({\bf k}) & d_x({\bf k})+id_y({\bf k}) \end{array}\right)
  \label{eq:dvector}
\end{eqnarray}
The orbital part is expressed by these {\bf d}-vector with
\begin{equation}
g_1=-d_x+id_y,~~~
 g_2=d_z, ~~~g_3=d_x+id_y,
\end{equation}
and the excitation energy is
\begin{equation}
E_{{\bf k}}=\sqrt{\xi_{\bf k}^2+\Delta^2|{\bf d}({\bf k})|^2}.
\end{equation}

In the presence of strong spin-orbit coupling only the total
angular momentum {\bf J=L+S} is a good quantum  number, and the
classification according to physical electron spin is not
possible.  However, if the crystal structure has inversion center,
Cooper pair states can still be classified according to their
parity, and therefore acquire a ``pseudospin'' quantum number,
instead of the physical spin. This situation is commonly
encountered in many heavy fermion materials.

Experimentally, the parity of the pair wave function
$\Psi_{pair}({\bf k})$ can be determined by the Knight shift of
the nuclear magnetic resonance (NMR) frequency
\cite{mineev,kitaoka}, muon spin rotation ($\mu$SR), and, less
directly, by the magnitude of the upper critical field $H_{c2}$.
The Knight shift is linear in the electron spin susceptibility
$\chi_s$, and is therefore a direct measure of the spin
polarization in the superconducting state.  In a spin singlet
superconductor, as Cooper pairs are formed, the spin contribution
to the Knight shift falls rapidly on cooling through the
transition. In contrast, in a triplet superconductor the spin
orientation of the Cooper pairs is determined by the {\bf
d}-vector in Eq.(\ref{eq:dvector}). If the direction $\bf d$ is
fixed by the spin-orbit interaction, the Knight shift is
anisotropic in the superconducting state. When the magnetic field
is applied along {\bf d} ({\bf H}$\parallel$ {\bf d}), the Cooper
pair spin is perpendicular to {\bf H} and hence does not
contribute to the susceptibility.  Then the Knight shift decreases
rapidly below $T_c$, as in spin singlet superconductors. On the
other hand, when the applied field {\bf H}$\perp$ {\bf d},  the
$x$- and $y$- components of {\bf d} vector, $d_x({\bf
k})+id_y({\bf k})$, are nonzero, and the contribution to the
susceptibility of the Cooper pairs is identical to that of the
constituent electrons, $\chi_{\parallel}=\chi_n$, where $n$ labels
the normal state above the transition temperature. Therefore the
Knight shift remains unchanged below $T_c$.

Applied magnetic field destroys superconductivity through both the
orbital dephasing and Zeeman splitting of the single electron
energy levels. In type-II superconductors the former effect leads
to the emergence of the vortex state, when the magnetic field
penetrates into the sample, forming the regular array of the
vortex tubes parallel to the field. Vortices have cores of the
size of the superconducting coherence length, $\xi$, where
superconductivity is destroyed.  Each vortex carries a flux
quantum $\Phi_0=\pi \hbar c/e=2\times 10^{-7}$G$\cdot$cm$^2$, and
therefore in the external field, $H$, the area per vortex, $A$, is
determined from $AH=\Phi_0$.  At the orbital upper critical field
the vortex cores overlap destroying bulk superconductivity, hence
a simple estimate gives $H_{c2}^O=\Phi_0/2\pi\xi^2$. On the other
hand, in singlet superconductors, an additional pairbreaking
effect of the field is due to polarization of the normal state
electrons. The upper critical field determined by this Pauli
limiting effect is estimated to be
$H_{c2}^P=\Delta/\sqrt{2}\mu_B$, where $\mu_B$ is the Bohr
magneton, and $\Delta$ is superconducting gap.  The Pauli limiting
is absent in spin triplet superconductors. Therefore, finding
$H_{c2}$ which is higher than $H_{c2}^P$ may indicate spin triplet
pairing.

Up to now, possible odd parity superconducting state has been
suggested in heavy fermion UPt$_3$\cite{UPt3NMR},
UNi$_2$Al$_3$\cite{UNiAlNMR}, URu$_2$Si$_2$\cite{URuSiK},
UBe$_{13}$\cite{ott}, PrOs$_4$Sb$_{12}$\cite{tou}, organic
(TMTSF)$_2$PF$_6$\cite{lee}, and transition metal oxides,
Sr$_2$RuO$_4$\cite{ishidaSr},
Sr$_2$Ca$_{12}$Cu$_{24}$O$_{41}$\cite{fujiwara} and Na$_x$CoO$_2
\cdot $yH$_2$O \cite{ishidaNa,yoshimura}.  It is most probably
realized in ferromagnetic superconductors UGe$_2$ \cite{UGe},
URhGe\cite{URhGe} and UIr\cite{akazawa}.   However, one needs to
bear in mind that the odd parity in some of these materials is
still controversial.  In fact,  analysis of the NMR spectrum in
the vortex state, from which we determine the Knight shift, is not
settled.    Moreover  the NMR experiment measures the surface area
with the length scale of penetration depth $\lambda$ .  In this
regime, strong currents associated with the surface barrier flow
and the field distribution and the susceptibility is strongly
space inhomogeneous.  The influence of these currents on the NMR
spectrum is an open question.  In addition, if the pairing
interaction is modified by magnetic field, the upper critical
field can be enhanced above $H_{c2}^P$ even in the spin singlet
superconductors.

When $\Psi_{pair}({\bf k})$ has an imaginary part
\begin{equation}
 \Psi_{pair}({\bf k})=\psi_1({\bf k})+i\psi_2({\bf k}),
\end{equation}
the time reversal symmetry is broken, since   $\Psi_{pair}^*({\bf
k}) \neq \Psi_{pair}({\bf k})$.  In such a situation, spontaneous
static magnetic field arising from the orbital current around the
impurity or at the surface can appear below $T_c$, because the
impurity or boundary lifts the degeneracy between  $|L,
L_z\rangle$ and  $|L, -L_z\rangle$.   Such a spontaneous magnetic
field was observed in UPt$_3$ \cite{mSRUPt3},
Sr$_2$RuO$_4$\cite{mSRSRO}, and PrOs$_4$Sb$_{12}$ \cite{mSRPr} by
$\mu$SR experiments.

Recently superconductivity with no spatial inversion symmetry has
excited great interest.    In the presence of strong spin-orbit
interaction, the absence of the inversion symmetry strongly
influence the pairing symmetry through a splitting of the two spin
degenerate bands.  Generally in the system without inversion
symmetry the gap function is a mixture of spin singlet and triplet
channels in the presence of a finite spin orbit coupling strength.
Associated with the absence of spatial inversion symmetry, unusual
superconduting properties, including the striking enhancement of
$H_{c2}$ and helical vortex phase, have been proposed.    The
absence of inversion symmetry has been reported in several
superconductors.  Among them,  the superconducting gap function
with line nodes have been reported in in
CePt$_3$Si\cite{izawaCePtSi} and Li$_2$Pt$_3$B \cite{yuan}.  In
these system, the position of node is strongly influenced by the
mixing of the spin singlet and triplet components.

In Table-I, we summarize the gap functions of several anisotropic
superconductors.   One needs to bear in mind that the gap
functions in some of the listed materials are still controversial.


\begin{sidewaystable}
\begin{center}

\caption{Superconducting gap symmetry of unconventional superconductors.  TRS, AFMO and FMO represent time reversal symmetry, antiferromagnetic ordering and ferromagnetic ordering, respectively }

\small{}
\begin{tabular}{|c|c|c|c|c|c|} \hline
                                                        & Node        &    Parity                             &    TRS     & Proposed gap function  & Comments      \\  \hline
high-$T_c$ cuprates                        &   line (vertical)       &     even \cite{tsuei}  &                &  $d_{x^2-y^2}$ \cite{tsuei}  &      \\  \hline
Sr$_2$Ca$_{12}$Cu$_{24}$O$_{41}$                       &   full gap   \cite{fujiwara}   &     odd \cite{fujiwara} &                &  &  spin ladder system  \\  \hline
$\kappa$-(ET)$_2$Cu(SCN)$_2$  &   line (vertical)  \cite{izawaET}         &   even \cite{maya}          &                &    $d_{xy}$ \cite{izawaET}   &    \\ \hline
(TMTSF)$_2$PF$_6$                     &       &   odd \cite{lee}                               &                 &       &    superconductivity under pressure\\
(TMTSF)$_2$ClO$_4$                     & line\cite{takigawa}      &                                &                 &       &    \\
                  &full gap \cite{BelinT}       &                                &                 &       &    \\ \hline
Sr$_2$RuO$_4$                              &line(horizontal)   \cite{izawaSr}         &    odd  \cite{ishidaSr}                         &broken \cite{mSRSRO}    &   $(k_x+ik_y)\cos(k_zc+\alpha)$ \cite{izawaSr}    &     \\
                            &line(vertical)   \cite{deguchi2}          &                             &     &      $\sin k_x+i\sin k_y $ \cite{deguchi2}&     \\ \hline
 Na$_x$CoO$_2 \cdot $yH$_2$O                                     &line \cite{fuji}         &even\cite{satoNa,zhangNa},                      &                 &  &\\
                                     &        &odd\cite{ishidaNa,yoshimura}                         &                 &  &\\ \hline
 (Y,Lu)Ni$_2$B$_2$C                             &  point-like   \cite{izawaYN2}       &   even   \cite{nohara}                    &           &  $1-\sin^4\theta\cos(4\phi)$ \cite{thalmaki,makithalwon}& very anisotropic $s$-wave \\ \hline
Li$_2$Pt$_3$B                             & line   \cite{yuan}       &   even +odd                      &           &  & no inversion center \\ \hline
CeCu$_2$Si$_2$                             & line  \cite{ueda}          & even  \cite{ueda}                               &                &  &two superconducting phases\cite{steg} \\ \hline
CeIn$_3$                                          &line\cite{kawasaki}           &                                 &                &  &coexistence with AFMO\\ \hline
CeCoIn$_5$                                     & line   (vertical)          & even \cite{izawaCe}                               &              & $d_{x^2-y^2}$\cite{izawaCe,WKPark2,Vorontsov1}, $d_{xy}$\cite{aoki}   &FFLO phase \\
CeRhIn$_5$                                      & line\cite{kohoriRh}         & even   \cite{kohoriRh}                                &               & &coexistence with AFMO \\ \hline
CePt$_3$Si                                      &line  \cite{izawaCePtSi}           &even+odd   \cite{bauer2}                      &                 &  &no inversion center \\ \hline
UPd$_2$Al$_3$                               & line(horizontal)& even \cite{touUPdAl}            &                  & $\cos k_zc$ \cite{watanabe} &coexistence with AFMO \\
UNi$_2$Al$_3$                                & line\cite{UNiAlNMR}             & odd\cite{UNiAlNMR}                               &                  & &coexistence with SDW \\ \hline
URu$_2$Si$_2$                               & line\cite{URuSiT1}             & odd \cite{URuSiK}                              &                  &  & coexistence with hidden order \\ \hline
UPt$_3$                                           & line+point \cite{Joynt}   & odd\cite{UPt3NMR}                               &broken\cite{mSRUPt3}        & & multiple superconducting phases  \\ \hline
UBe$_{13}$                                     & line  \cite{Mac}            & odd \cite{tou}                          &                    & & \\ \hline
UGe$_{2}$                                      & line \cite{kote}             & odd  \cite{UGe}                             &                    &  &coexistence with FMO\\ \hline
URhGe                                             &                & odd   \cite{URhGe}                             &                     &  &coexistence with FMO\\ \hline
UIr                                           &                & even+odd  \cite{akazawa}                            &                     &  &coexistence with FMO\\
                                          &                &                                &                     &  &\&no inversion center\\ \hline
PuCoGa$_5$                                    &line    \cite{curroPu}         & even  \cite{curro}                            &                  & &\\
PuRhGa$_5$                                    &line \cite{sakai}             &                               &                  & &\\ \hline
PrOs$_4$Sb$_{12}$                        &point   \cite{izawaPr}          & odd \cite{tou}                               &broken\cite{mSRPr}        &  &multiple superconducting phases\\ \hline

\end{tabular}

\normalsize{}

\end{center}
\end{sidewaystable}

\section{Nodal Structure: Standard Techniques}

In  unconventional superconductors discovered so far,
the  energy gap for the quasiparticle excitations vanishes (has
nodes) at points or along lines on the Fermi surface. There is now
a wide variety of dynamic and thermodynamic probes that couple to,
and reveal these low energy excitations. The temperature
dependence of the London penetration depth $\lambda(T)$,
electronic part of the specific heat $C(T)$, thermal conductivity
$\kappa(T)$, and nuclear magnetic resonance (NMR) spin-lattice
relaxation rate $T_1^{-1}$ all reflect the changes in the
quasiparticle occupation numbers. In the fully gapped ($s$-wave)
superconductors the quasiparticle density of states (DOS) is zero
at energies, $E$, below the gap edge (no excitations with
$E<\Delta$), and  varies as
$N_s(E)/N_0=\frac{E}{\sqrt{E^2-\Delta^2}} $ for $E>\Delta$. Here
$N_0$ is the normal state DOS. The physical quantities exhibit
activated temperature dependence, $\exp(-\Delta/T)$ at low
temperatures, $T\ll T_c$. On the other hand, in nodal
superconductors, the low-energy density of states remains finite
due to contributions from the near-nodal regions on the Fermi
surface, and typically the DOS varies as
$N_s(E)/N_0=(E/\Delta_0)^n$ at $E\ll\Delta$. The exponent $n$
depends on the topology of the nodes: $n=1$ for line nodes as well
as for point nodes where the gap is quadratic in distance from the
nodal point in the momentum space; $n=2$ for point nodes with
linearly varying gap amplitude around the nodal point. Then the
experimental quantities described above exhibit power law
temperature dependence at $T\ll T_c$. For example, in $d$-wave
superconductors with line nodes, the DOS
$N(E)\sim|E|$ leads to the specific heat $C_e\approx\gamma_n
T^2/T_c$, where $\gamma_n$ is the coefficient of the
linear-$T$-term in the normal state, and the NMR relaxation rate $T_1^{-1}  \propto T^3$. The deviation of the superfluid density, $n_s(T)$
from its zero temperature value,
$\Delta n_s(T)=n_s(0)-n_s(T) \propto k_BT/\Delta_0$, which can be detected by the penetration depth measurements.


So far we discussed pure systems. The regime where power laws in
$T$ are observed is strongly influenced by the impurities.  In
unconventional superconductors non-magnetic impurities act as
pair-breakers, similar to magnetic impurities in $s$-wave
superconductors.  A bound state appears near an isolated
non-magnetic strong (scattering phase shift $\pi/2$, or unitarity)
scatterer, at the energy close to the Fermi level. The broadening
of this bound state to an impurity band at finite disorder leads
to a finite density of states at zero energy, $N(0)$, that
increases with increasing impurity concentration \cite{hirsch}.
The impurity scattering changes the temperature dependence of the
physical quantities below $T$ corresponding to the impurity
bandwidth: $\Delta\lambda$ changes the behavior from $T$ to $T^2$,
the NMR relaxation rate changes from $T_1^{-1} \propto T^3$ to
$T_1^{-1} \propto T$ ($T_1T\sim const.$), and $C(T)$ changes from
$T^2$ to $T$. In NMR, for example,  the temperature range where
$T_1^{-1}$ exhibits the $T^3$- dependence is limited to $T>T_c/3$
in most measurements.

Therefore the nodal behavior may be hidden by the impurity
effects. Even in the extremely pure systems, however, experimental
observation of the power laws provides indications of the
existence of the nodes, but is unable to yield information about
their location in the momentum space.

Angle resolved photoemission spectroscopy (ARPES) directly
investigates the momentum dependence of the gap, and was
instrumental in determining the gap shape in high-T$_c$
superconductors. However, the energy resolution of this technique
is insufficient when compared with the size of the energy gap in
most low-$T_c$ systems.  The phase-sensitive measurements, such as
corner junctions, tricrystal, and tests for Andreev bound states,
which provided the most convincing evidence for $d_{x^2-y^2}$ wave
order parameter in the high-$T_c$ cuprates, are primarily surface
probes. Absence of inversion symmetry near the surface may
influence the pairing symmetry through, for example, splitting of
the two spin degenerate bands via the spin-orbit coupling, so that
the gap function is a mixture of the spin-singlet and spin-triplet
channels. Moreover, it is difficult to apply these techniques to
determine the three dimensional gap structure, and they have
received limited use beyond studies of the high-$T_c$ cuprates. It
is therefore extremely important to acquire complementary evidence
for particular gap symmetries via bulk measurements.

\section{Nodal Superconductor in Magnetic Field}

\subsection{General approaches to the vortex state.}

Determining the nodal positions requires a directional probe. In
the following we argue that an applied magnetic field provides a
convenient bulk probe of the symmetry of the order parameter in
unconventional superconductors. The usefulness of the magnetic
field as a probe relies on an important difference between the
properties of the vortex state in nodal compared to fully gapped
$s$-wave superconductors. While for the $s$-wave case the DOS and
the entropy at low fields, $H\ll H_{c2}$, are determined by the
localized states in the vortex cores, in the superconductors with
nodes they are dominated by the extended quasiparticle states,
which exist in the bulk close to the nodal directions in momentum
space. Therefore much attention has been paid to the effect of the
field on these near-nodal quasiparticle.

A simple picture that captures the main effect of the magnetic
field on a nodal superconductor is that of the Doppler shift of
the quasiparticle spectrum \cite{volovik}. In the presence of a
supercurrent with velocity {\bf v}$_s$ the energy of a
quasiparticle with momentum {\bf k} is Doppler shifted relative to
the superconducting condensate by
\begin{equation}
\varepsilon({\bf k}) \rightarrow \varepsilon({\bf k})-\hbar{\bf k}\cdot{\bf v}_s.
\end{equation}
This effect originates from the Galilean transformation:  creation
of an excitation $({\bf p}, \varepsilon)$ in the
normal-quasiparticle rest frame, involves an additional energy
$\delta\varepsilon={\bf p}\cdot{\bf v}_s$ in the superfluid frame
of reference. For a uniform superflow the Doppler shift is simply
a consequence of the gauge invariance, and is therefore exact
\cite{MTinkham}. For a non-uniform superflow, as in the vortex
state, this picture is semiclassical in that it considers
simultaneously the momentum of the quasiparticles and the local
value of ${\bf v}_s(\bf r)$ at position $\bf r$, and therefore
ignores the possible accumulation of the quantum mechanical phase
around the magnetic vortices in superconductors. The fully quantum
mechanical treatment of the quasiparticle energies so far  was
carried out only in a perfectly periodic vortex lattice and in the
absence of impurities \cite{MFranz:sing}, and gives results for
the physical properties close to those obtained in the
semiclassical treatment \cite{MFranz:sing,DKnapp}.

To estimate the characteristic energy scale of the Doppler shift
we can approximate the velocity field by that around a single
vortex, ${\bf v}_s=\hbar \widehat{\phi}/2mr$, where $r$ is the
distance from the center of the vortex and $\widehat{\phi}$ is a
unit vector along the circulating current. This expression is
valid outside the vortex core and up to a cutoff of order
min\{$R,\lambda$\}, where $R=a\sqrt{\Phi_0/\pi H}$ is the
intervortex distance, $\Phi_0$ is the flux quantm, $a$ is a
geometric constant, and $\lambda$ is the London penetration depth.
Average Doppler shift, $E_{av}$, is computed by integrating over a
vortex lattice unit cell, and is given by
\begin{equation}
E_{av}=\langle |{\bf v}_s\cdot {\bf p}| \rangle =\int_{|\bf r|<R}
\frac{d^2\bf r}{\pi R^2} |{\bf p}\cdot{\bf v}_s|
 \approx \frac{4}{a \pi}\hbar v_F\sqrt{\frac{H}{\Phi_0}}.
\end{equation}
Thus $E_{av}$ is proportional to $\sqrt{H}$.

Since the density of states is an additive quantity, the net DOS
of the sample is the sum of the contributions from the areas with
distinct values of the Doppler shift. In a system with line nodes,
where the low energy DOS $N(E)\propto |E|$, this implies the
residual density of states $N_v(E=0,H)\propto E_{av}\propto
\sqrt{H}$. Consequently the specific heat also exhibits the
$\sqrt{H}$-behavior at low temperatures in the clean limit
\cite{volovik,kubert1,moler1,moler2}. Since the supervelocity
distribution can be obtained for a given configuration of
vortices, the range of possible values for the coefficient of the
$\sqrt H$ can also be found for a given material \cite{vekhter3}.

Determining the transport properties, such as the thermal
conductivity $\kappa$, using the Doppler shift method is a more
challenging task. The transport coefficients are determined from
the correlation functions that have a finite range, and therefore
depend on the Doppler shift at more than a single point. Local
values of these coefficients can be rigorously defined only in the
dirty limit. It is generally accepted that a similar definition
gives at least a qualitatively correct results in the clean limit
\cite{kubert}, although a rigorous comparison is currently
lacking. Even with that assumption, the connection between the
distribution of local values, $\kappa(\bf r)$, and the measured
value remains a subject of some debate. Both averaging $\kappa(\bf
r)$ and $\kappa^{-1}(\bf r)$ have been proposed
\cite{kubert,maki1}. While in some cases the difference is only in
the magnitude of the field-induced change, the divergent
philosophies behind the averaging procedures give rise to
qualitatively different results for the anisotropy of the
transport coefficients described below.

In the approach discussed above only the quasiparticle energy is
shifted, so that the single particle scattering rate is not
directly affected by the presence of the vortices. In the presence
of static disorder treated, for example, in the self-consistent
$T$-matrix approximation, the magnetic field does affect the
lifetime indirectly, by modifying the density of states available
for scattering \cite{kubert1,kubert}. Hence the Doppler shift
method does not account for scattering of the quasiparticles on
vortices.

Calculations of the vortex scattering cross-section have to go
beyond the semiclassical treatment and make assumptions about the
structure of the vortex core states \cite{Durst}, and therefore
received limited attention. An alternative, fully microscopic
approach, employs an extension of the approximation originally due
to Brandt, Pesch and Tewordt (BPT) \cite{BPT,BPT1}  to describe
clean superconductors near the upper critical field, $H_{c2}$. In
this method the Gor'kov equations (or their quasiclassical
Eilenberger-Larkin-Ovchinnikov analog) are solved with the normal
electron Green's function replaced by its spatial average
\cite{BPT1,vekhter2,houghton1,kusunose04,Vorontsov1}. This method
is rigorously justified at moderate to high fields, gives the
standard quasiparticle spectrum as $H\rightarrow 0$, and yields
results that are qualitatively similar to the Doppler shift at
very low fields and temperatures. Therefore it is believed that it
can be used over a wide range of fields relevant to experiment.
The BPT approach naturally includes the scattering of the
quasiparticles off the vortices \cite{vekhter2,Vorontsov1};
however, due to the incoherent averaging over different unit cells
of the vortex lattice, it tends to overestimate the importance of
such scattering at lower fields. Together, BPT and Doppler shift
methods account for a majority of theoretical work relevant to the
experimental investigations of the quasiparticle properties in the
vortex state of nodal superconductors.

\subsection{Thermal Conductivity}

\begin{figure}[t]
\begin{center}
\includegraphics[width=9cm]{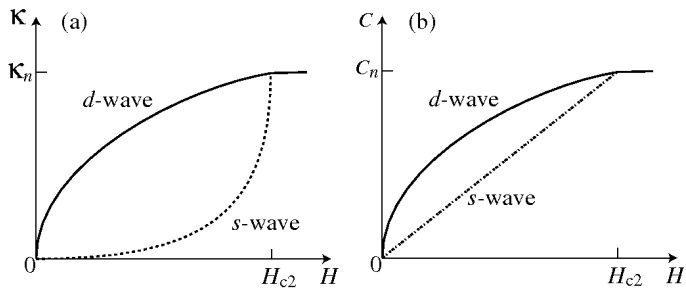}
\caption{Field dependence of  (a) the thermal conductivity $\kappa$ and
(b) the specific heat $C$ for $s$- and $d$-wave superconductors.  In $s$-wave superconductors, the thermal conductivity shows an exponential behavior with very slow growth with $H$ \cite{Nb,KasaharaK}.  The heat capacity increases nearly linearly with $H$.   In sharp contrast, in  $d$-wave superconductors,
both the specific heat and the quasiparticle conduction grows rapidly as soon as the field exceeds
 $H_{c1}$. The slope of $\kappa(H)$ at $H_{c2}$ depends on purity and therefore for the $d$-wave case it is also possible to
 have an inflection point in $\kappa(H)$ at intermediate fields.}
\label{fig:H_dep_kappa_C}
\end{center}
\end{figure}

Of all the transport properties the thermal conductivity is
uniquely suitable for probing bulk superconductivity.  Unlike
electrical resistivity, it does not vanish in the superconducting
state. Cooper pairs do not carry entropy and therefore do not
contribute to the thermal transport. As a result, the thermal
conductivity probes the delocalized low energy quasiparticle
excitations, and  is sensitive to the effect of magnetic field on
the quasiparticles.  In Fig.~\ref{fig:H_dep_kappa_C} we show the
qualitative behavior of the thermal conductivity and heat capacity
at low $T$ as a function of the magnetic field for a  $s$-wave (fully gapped) and a $d$-wave (with line
nodes) superconductor.

In  $s$-wave superconductors  the only quasiparticle states
present at $T \ll T_c$ are those associated with vortices.  At low
fields where the vortices are far apart, these states are bound to
the vortex core and are therefore localized and unable to
transport heat; the thermal conductivity shows an exponential
behavior with very slow growth with $H$.  At high fields near
$H_{c2}$ where quasiparticles states within the vortices begins to
overlap with those within the neighboring vortices,  thermal
conductivity increases rapidly.  Such a field dependence of the
thermal conductivity is observed in Nb \cite{Nb,KasaharaK}.  The
heat capacity, due to the localized quasiparticle states,
increases nearly linearly with $H$.   In dramatic contrast, both
the specific heat and the quasiparticle conduction, due to near
nodal states, grow rapidly as soon as the field exceeds $H_{c1}$.
In $d$-wave superconductors where $N(E=0,H)\propto \sqrt{H}$ due
to the Doppler shift of the quasiparticle energy spectrum,   both
the thermal conductivity and the specific heat exhibit a nearly
$\sqrt{H}$-behavior.

In reality, especially at higher temperatures, the behavior of the
thermal conductivity is more complex. While the magnetic field
enhances the local DOS, it also leads to a change in the transport
lifetime both via the modification of the impurity scattering and
via Andreev scattering off the vortices. Understanding of these
competing effects has progressed during the past few years
\cite{vekhter3,kubert,vekhter2,Vorontsov1,barash,franz,yu,aubin},
although the complete picture is not yet developed. In general, at
low temperatures the DOS modification plays the dominant role, and
the thermal conductivity increases with increased field. At higher
temperatures and low fields, the dominant effect of vortices is to
introduce and additional scattering mechanism, while the DOS is
controlled by $T$. Consequently, the thermal conductivity
initially decreases with field, and goes through a minimum at a
finite $H$ \cite{vekhter2}. This behavior has been first observed
in high-T$_c$ cuprates \cite{Ong}, and also seen in other systems
\cite{izawaCe,watanabe}

\section{How to determine the nodal structure in the bulk?}

\subsection{Anisotropy under rotating field: density of states}

The techniques described above help determine  the general
topology of the gap,  but cannot establish the exact angular
dependence of $\Delta$({\bf k}). In particular, the nodal
positions in {\bf k}-space cannot be obtained. In the
following we discuss the theoretical underpinnings and
experimental realizations of the new and powerful method for
determining the nodal directions.

The method is based on the prediction that, under an applied
magnetic field, the density of states of nodal superconductors
depends on the orientation of the field with respect to the nodal
direction \cite{vekhter}, and exhibits characteristic oscillations
as the field is rotated relative to the sample. The oscillations
can be measured via the field-angle dependence of the thermal
conductivity
\cite{yu,aubin,izawaCe,watanabe,ocana,izawaSr,izawaET,izawaYN2,izawaPr}
or specific heat \cite{aoki,park1,park2,deguchi1,deguchi2}.

While the specific heat measurements directly probe the density of
states, the thermal conductivity anisotropy is sensitive to a
combination of the density of states, and the quasiparticle
transport scattering rate, which may have different dependence on
the field orientation. Measurements of the specific heat
anisotropy were only attempted several years after theoretical
predictions \cite{aoki,park1,park2,deguchi1,deguchi2}. First
experiments on the field-angle dependence of the thermal
conductivity preceded theoretical discussions \cite{yu,aubin}, but
focused simply on the existence, rather than location, of
additional features (interpreted as arising from Andreev
scattering) in cuprates. Use of the thermal conductivity as a
similar test based on the DOS anisotropy  was, to our knowledge,
first suggested in Ref.\cite{Hirschfeld-Korea}, and followed up by
other work \cite{maki1}. Development and consistent use of the
measurements to probe the direction and type of nodes in
{\bf k}-space, is almost entirely due to recent efforts by
the group of University of Tokyo
\cite{izawaCe,watanabe,izawaSr,izawaET,izawaYN2,izawaPr}. The full
theory of the anisotropy of the thermal conductivity in the vortex
state is still incomplete. Hence, while the salient features of
experiments are qualitatively understood based on a number of
treatments
\cite{maki1,Vorontsov1,vekhter,nakai1,nakai2,kusunose1,maki2,udagawa1,udagawa2,kusunose2,thalUPdAl1,adachi},
many details need to be addressed further. Below we discuss the
current status of this field.

\begin{figure}[t]
\begin{center}
\includegraphics[width=9cm]{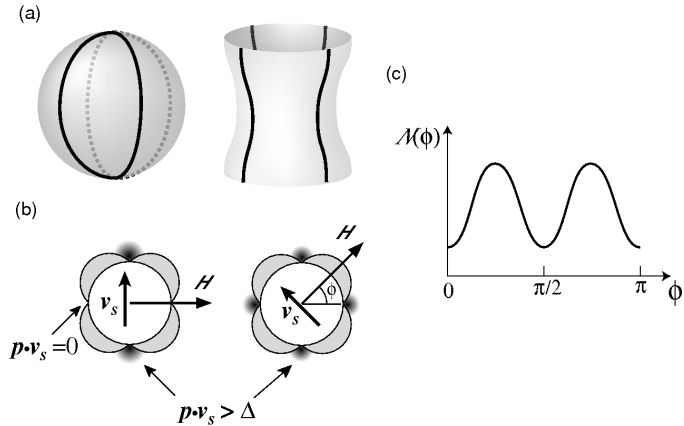}
\caption{(a)Sketch of the gap structure with four line nodes
perpendicular to the basal plane (vertical node). (b)Schematic
diagram showing the regions on the Fermi surface that experience
the Doppler shift in {\boldmath $H$} within the basal plane.  We
have assumed $d_{xy}$ symmetry.  $\phi$ = ({\boldmath $H$},$a$) is
the azimuthal angle measured from the $a$-axis.   With {\boldmath
$H$} applied along the antinodal directions, all four nodes
contribute to the DOS, while for {\boldmath $H$} applied parallel
to the node directions, the Doppler shift vanishes at two of the
nodes. (c) Four-fold oscillation of the DOS for {\boldmath $H$}
rotating in the basal plane.  The DOS shows a maximum (minimum)
when {\boldmath $H$} is applied in the anitinodal (nodal)
direction.} \label{fig:Doppler_shift_v}
\end{center}
\end{figure}

The origin of the anisotropy is best understood in the framework
of the Doppler shift of the delocalized quasiparticle spectrum in
the vortex state. Consider, for simplicity, a $d_{xy}$ gap
symmetry with four vertical lines of nodes, and assume a
cylindrical or spherical  Fermi surface, as illustrated in
Figs.\ref{fig:Doppler_shift_v} (a)-(c). At low fields, the loci of
unpaired quasiparticles in the momentum space are close to the
nodal lines. Since the supercurrents flow in the plane normal to
direction of the applied field, the Doppler shift experienced by
quasiparticles in a given near-nodal region depends on the
direction of the field {\boldmath
$H$}$(\theta,\phi)=H(\sin\theta\cos\phi, \sin\theta\sin\phi,
\cos\theta)$ with respect to the nodal directions. Here  $\theta$
and $\phi$ are  the polar angle and the azimuthal angle
respectively, measured relative to the $c$-axis.

Consider  {\boldmath $H$} rotated  conically (fixed $\theta$) with
varying in-plane angle, $\phi$; see the view from above in
Fig.~\ref{fig:Doppler_shift_v}(b). When the field is aligned with
a nodal line, the superflow around the vortices is in the plane
nearly normal to the momenta of quasiparticles close to that node.
As a result, for these quasiparticles the Doppler shift is small.
In contrast, when the field is in the antinodal direction, the
Doppler shift is (relatively) large along all four nodal lines. As
a result, the net DOS has minima when {\boldmath  $H$} is aligned with the
nodal direction, and maxima for {\boldmath  $H$} along the antinodes
\cite{vekhter}. The angle-dependence of the DOS exhibits
characteristic four-fold oscillations, as shown in
Fig.~\ref{fig:Doppler_shift_v}(c). In general, DOS oscillates with
$n$-fold symmetry corresponding to the number of vertical nodes
$n$.

In this approach the amplitude of the DOS oscillations, $\delta
N(E)/N(E)$, depends on the shape of the Fermi surface and other
parameters of the models. For the residual ($E=0$) DOS, most
calculations predict the oscillation amplitude ranging from 3\% to
10\%. The anisotropy is rapidly washed away at finite energy, and
therefore the amplitude of the corresponding oscillations in the
measured quantities, such as the specific heat, at finite
temperature, is typically of the order of a few percent.

\begin{figure}[t]
\begin{center}
\includegraphics[width=12cm]{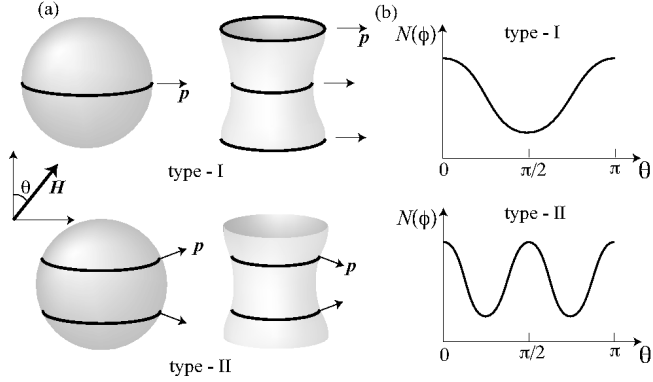}
\caption{ (a) Schematic figure of the gap structure with line nodes parallel to the basal plane (horizontal node) in spherical and open Fermi surfaces.  Type-I assumes a gap function $\Delta(\mbox{\bf k}) \propto \sin k_zc$, where line nodes are located at the center of the Brillouin zone and at the zone boundary.  Type-II assumes a gap function $\Delta(\mbox{\bf k}) \propto \cos k_zc$, where line nodes are located at positions shifted off the zone center.   (b) Oscillations of the DOS for {\boldmath $H$} rotating in the $ac$-plane for various gap functions.  Two-fold oscillation with the same sign are expected for type-I.  On the other hand, for type-II, an oscillation with a double minimum is expected.}
\label{fig:Doppler_shift_h1}
\end{center}
\end{figure}

Consider now horizontal line nodes in a cylindrical or spherical Fermi surface,
as illustrated  in Fig.\ref{fig:Doppler_shift_h1} (a).
The density of states is anisotropic under the rotation
of the field in the $ac$-plane, by varying the angle $\theta$.
To illustrate the difference between the line nodes at high
symmetry positions in the Brillouin zone, and away from those, we
consider here two model gap functions
\begin{enumerate}
\item type-I : Horizontal nodes located at the center of the Brillouin zone and at the zone boundary
$\Delta(\mbox{\bf k}) \propto \sin k_zc$.
\item type-II : Horizontal node located at positions shifted off the zone center;
$\Delta(\mbox{\bf k}) \propto \cos k_zc$.
\end{enumerate}
The expected angular variation of the Doppler shifted DOS is a
function of the relative angle between {\boldmath $H$} and {\bf p}
for these gap functions is shown schematically in
Fig.\ref{fig:Doppler_shift_h1} (b).   The twofold oscillation is
expected for type-I gap functions, in which the horizontal nodes
are located at the position where {\bf p}$\parallel ab$-plane.  On
the other hand, for type-II, one expects an oscillation with a
double minimum structure as a function of $\theta$. Note that we
sketched the DOS for a fixed $H/H_{c2}$; if a measurement is done
at a fixed $H$, the anisotropy of $H_{c2}$ in a quasi-2D systems
superimposes an additional two fold component on the oscillations,
so that for type-II gap the central maximum is distinct from the
other two.

The Doppler shift method does not account for the scattering of
electrons on vortices. Fully microscopic analyses  indicate that
inclusion of such scattering further reduces the amplitude of the
DOS anisotropy \cite{vekhter3}. Furthermore, it has been shown
very recently that at moderate to high fields and temperatures the
vortex scattering leads to the inversion of the anisotropy: the
density of states is greater for the field along the nodal
directions than for the field along the gap maxima
\cite{Vorontsov1}. While this makes the analysis of the specific
heat data more complicated, it affects the conclusions drawn from
the analysis of the transport properties less dramatically
\cite{Vorontsov1}, as we discuss in next section.

\subsection{Anisotropy under rotating field: thermal conductivity}

Our focus in this review is on the determination of the nodal
structure via the thermal conductivity measurements. The
anisotropy of transport coefficients is given by a combined effect
of the angular variations of the density of states, and the
angle-dependent scattering. The latter effect is not yet fully
understood. Since in self-consistent treatments the scattering
rate of quasiparticles off impurities depends on the density of
states, the impurity scattering under a rotating field acquires
the same $n$-fold anisotropy as the DOS. However, depending on the
strength of impurity scattering, temperature, and the field, the
lifetime may exhibit either maxima or minima for the field aligned
with the nodes. Consequently, the possibility of the inversion of
the maxima and minima in the $T-H$ superconducting phase diagram
(which was unexpected in the behavior of the DOS and the specific
heat) was anticipated in the anisotropy of the transport
coefficients.

It is believed that the field dependence of the thermal
conductivity indicates whether the lifetime or the density of
states effects dominate. In the regime where $\kappa(H)$ decreases
with increasing field due to field-enhanced scattering, the maxima
of the anisotropic conductivity are likely to correspond to nodal
direction. In contrast, when $\kappa(H)$ increases with field, the
density of states effects dominate, and the minima of the $n$-fold
pattern indicate the nodes. This conjecture is not rigorous
\cite{Vorontsov1}, but qualitatively correct and provides guidance
in the situations when no results of microscopic theory are
available for a given compound. Moreover, in experiment the
angle-induced anisotropy of $\kappa(H)$ changes sign close to the
point where the field dependence has a minimum, supporting this
view.

When heat current, $\bf j_h$, is applied in the basal plane, the
angle between $\bf j_h$ and $\bf H$ is varied as the field is
rotated. Consequently, the dominant anisotropy observed in
experiment is that between the transport along and normal to the
vortices, i.e. twofold \cite{maki67}. The nodal contribution
appears as a smaller effect on this background, as was first seen
in the high-$T_c$ cuprate YBa$_2$Cu$_3$O$_{7-\delta}$
\cite{yu,aubin,ocana}.  Note that with few exceptions
\cite{HirschfeldC} Doppler shift does not describe the combined
twofold and nodal anisotropy.

More sophisticated approaches based on the BPT theory give correct
shapes of the $\kappa(\phi)$ curves, and account for most of the
observed features. The details of the competition between the
twofold and the fourfold oscillations depend on the shape of the
Fermi surface, role of Zeeman splitting, impurity strength and
concentration etc. Therefore any semi-quantitative comparison of
theory and experiment requires knowledge of these as an input, and
has only been done for few systems. At the same time qualitative
conclusions about the shape of the gap can still be drawn from the
simplified analysis, and we review those for the specific
compounds discussed below.

For relatively three-dimensional systems,  the current can be
applied along the $c$-axis, and the field rotated conically,
varying the azimuthal angle  $\phi$, and keeping the polar angle
$\theta$ constant. In that case the relative orientation of the
heat current and the field remains unchanged, and the oscillations
reflect solely the nodal structure. Vortex scattering still
modifies the amplitude and the sign of these oscillations, but the
interpretation is greatly simplified by the absence of the
dominant twofold term.

\begin{figure}[t]
\begin{center}
\includegraphics[width=10cm]{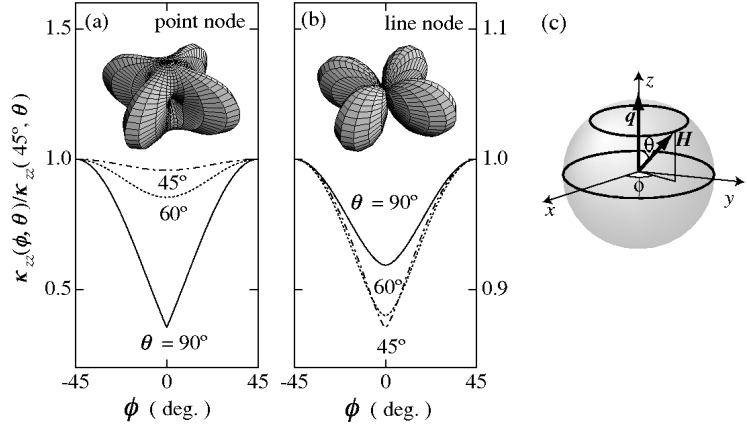}
\caption{ The thermal conductivity $\kappa_{zz}(\phi,\theta)$ (the
heat current {\bf q} $\parallel z$) when the magnetic
field is rotated conically as a function of $\phi$, keeping
$\theta$ constant,  for two different types of nodes,  (a)point
and (b) line. $\kappa_{zz}(\phi,\theta)$ is normalized by
$\kappa_{zz}(45^{\circ}, \theta)$. The corresponding gap functions
are illustrated in the insets. (c) Definition of $\theta$ (polar
angle) and $\phi$ (azimuthal angle).  After Ref.\cite{thalmaki}. Note that for
vertical line nodes the relative amplitude of oscillations for
$\theta$ between 45$^\circ$ and 90$^\circ$ depends on the shape of
the Fermi surface, not accounted for in Ref.{thalmaki}}
\label{fig:Line_point}
\end{center}
\end{figure}

In this geometry it has been predicted that the
$\theta$-dependence of the  shape and the amplitude of the
periodic oscillations provide direct information on the type of
nodes, point or line \cite{izawaYN2}. In Figs.
\ref{fig:Line_point} (a) and (b), we compare the angular variation
of the thermal conductivity $\kappa_{zz}$  (the heat current
{\bf q} $\parallel z$) when the magnetic field is rotated
conically as a function of $\phi$, keeping $\theta$ constant,  for
two different types of nodes calculated from the Doppler-shifted
QP spectrum, in accordance with Ref.\cite{izawaYN2}.  Here we
adopted  gap functions $\Delta(\mbox{\bf
k})=\Delta_0\sin(2\phi)$ ($d$-wave)  for line node, and
$\Delta(\mbox{\bf k})=\frac{1}{2}\Delta_0
\{1-\sin^4\theta\cos(4\phi)\}$ for point node. The latter was
proposed in Ref.\cite{maki3}, but is probably not realized in this
system, and we use it as a convenient ansatz to illustrate the
behavior due to point nodes. These gap functions are illustrated
in the insets of Figs.\ref{fig:Line_point} (a) and (b).    Here
the clean limit $\frac{\hbar\Gamma}{\Delta}\ll\frac{H}{H_{c2}}$ is
assumed, where $\Gamma$ is the carrier scattering rate.

According to the Doppler shift picture, there are two major
differences in the angular variation of the thermal conductivity
between point nodes and line nodes. First, the shape of the
$\kappa(\phi)$ curves is different. While the oscillation is close
to a sinusoidal wave for line node (Fig.~\ref{fig:Line_point}(b)),
a narrow cusp structure is predicted for the point node at $T=0$
(Fig.~\ref{fig:Line_point}(a)). Qualitatively, the cusp appears as
a result of the small phase space available for the quasiparticles
induced in the vicinity of point nodes by the applied field.  For
line node the corresponding phase space is greater, and the
minimum is not as sharp. Second, the amplitude of the oscillation
at $T=0$ decreases rapidly when the {\boldmath $H$} is rotated
conically as a function of $\phi$ keeping $\theta$ constant. For
point nodes, the amplitude of the oscillation of the thermal
conductivity at $\theta=45^{\circ}$ is much smaller than that at
$\theta=90^{\circ}$, while they are of almost the same magnitude
for line nodes.  This can be accounted for considering the fact
that  for  $\theta=45^{\circ}$ geometry the field {\boldmath $H$}
is never aligned  with the point nodes on the equator. Hence there
is always a finite Doppler shift at all the nodes. In contrast,
for vertical line nodes, the rotating {\boldmath $H$} at any
$\theta$ always crosses the line of nodes leading to a greater
suppression of the DOS.

While no microscopic calculations exit at present for this
geometry, it is likely that the main conclusions of the Doppler
shift picture remain valid. As discussed above, the salient
features of the measurement in the conical experimental geometry
are less sensitive to the vortex scattering than those measured
with the heat current in the basal plane. It is likely that the
sharp cusp is smeared by finite temperature, but the rapid decay
of the oscillations as the field is tilted away from the plane
must remain observable. Nonetheless, more work utilizing
microscopic theory is clearly desirable in this situation.

In summary, the variation of the field direction in
$(\theta,\phi)$ leads to periodic variations in both the thermal
conductivity and the heat capacity of nodal superconductors.  From
the periodicity, phase and shape of the angular variation of the
thermal conductivity and heat capacity, one can extract
information on the direction and type of nodes in {\bf
k}-space.

 \subsection{Experimental}

In the experiments described below the thermal conductivity was
measured in a $^3$He cryostat by using the standard steady-state
method, with a heater and two carefully calibrated RuO$_2$
thermometers.   In all the measurements, and especially in
quasi-2D superconductors with very anisotropic upper critical
field, it is critically important to align {\boldmath $H$} in the plane
with high accuracy, and have a good control over its rotation.
Even a slight field-misalignment may produce a large effect on the
measured $\kappa$, influencing the conclusions. To achieve this
high precision for the orientation of {\boldmath $H$} relative to the
crystal axes, we used a system with two superconducting magnets
generating magnetic fields in two mutually orthogonal directions,
and a $^3$He cryostat set on a mechanical rotating stage at the
top of a Dewar. By computer-controlling the two superconducting
magnets and rotating stage, we were able to rotate {\boldmath $H$}
with a misalignment of less than 0.02$^{\circ}$ from each axis,
which we confirmed by simultaneous measurements of the
resistivity.

Since the thermal conductivity can be measured both under the
field {\boldmath $H$}=$H(\sin\theta\cos\phi, \sin\theta\sin\phi,
\cos\theta)$ rotated within the basal $ab$-plane (as a function of
$\phi$ at $\theta=90$), and {\boldmath $H$} rotated as a function
of $\theta$ at fixed $\phi$,  we were able to detect both vertical
and horizontal nodal structure.   In addition, measuring the
thermal conductivity with {\boldmath $H$}=$H(\sin\theta\cos\phi,
\sin\theta\sin\phi, \cos\theta)$ rotated conically as a function
of $\phi$, keeping $\theta$ constant, as shown in
Fig.\ref{fig:Line_point}, enables us, at least in principle,  to
distinguish line and point nodes. In the following we discuss the
experimental  results for different compounds.

One of the recurring aspects of the discussion is the relative
importance of electron and phonon (or spin-wave) contributions to
the net thermal conductivity. While at low temperatures the bosons
are less efficient than fermions in carrying heat, in systems with
significant spin fluctuations or low carrier density the bosonic
degrees of freedom may be dominant over a wide $T$-$H$ range.
Since only electrons carry charge current, we take the point of
view that if the measured Wiedemann-Franz ratio of the thermal to
electrical conductivities, $L = \kappa_{zz}\rho_{zz}/T$, just
above $T_{c}$ is close to the Lorenz number $L_{0}= 2.44\times
10^{-8}$~$\Omega$W/K, obtained under assumption of purely
electronic $\kappa$, the electronic contribution to the thermal
conductivity is dominant. Even though, strictly speaking, in the
presence of inelastic scattering, $L\rightarrow L_0$ only as
$T\rightarrow 0$, quite generally opening of an additional bosonic
conduction channel increases $L$ compared to $L_0$, and has clear
experimental signatures.

In several of the systems we discuss  $\kappa(T)/T
>\kappa(T_c)/T_c$ at least in some range $T\lesssim T_c$ and often has  peak at $T^\star<T_c$.
In compounds with low carrier density, and therefore with thermal
transport dominated by bosonic degrees of freedom, this may be due
to increased mean free path of phonons as the unpaired electron
density is decreased. In other materials, such increase is due to
rapid reduction of the inelastic scattering rate below $T_c$
(faster than the concomitant reduction in the electron density of
states). In some systems, both effects combine. Generally, a
comprehensive analysis of a large body of data on a given compound
provides a clue to what mechanism is more important. We indicate
this for each of the materials analysed below.

\section{Three Dimensional Unconventional superconductors}

 \subsection{Borocarbide YNi$_2$B$_2$C}

\begin{figure}[b]
\begin{center}
\includegraphics[width=5cm]{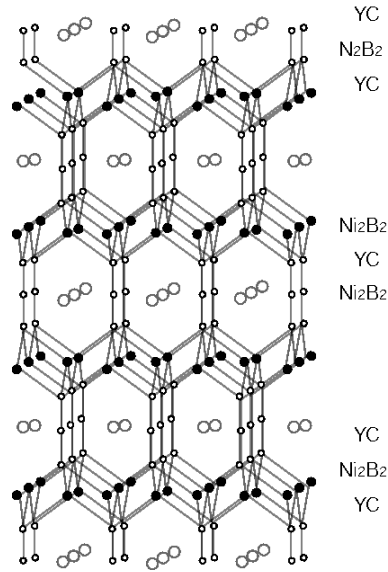}
\caption{Crystal structure of YNi$_2$B$_2$C.}
\label{fig:YNi2B2C_cr}
\end{center}
\end{figure}

We start by considering the superconducting  gap structure of a
non-magnetic borocarbide superconductors LnNi$_2$B$_2$C, Ln=(Y and
Lu)\cite{canfield}. These systems have tetragonal crystal
symmetry, and the electronic band structure is essentially 3D, see
Fig.\ref{fig:YNi2B2C_cr}.  Early on, these materials were assumed
to have an isotropic $s$-wave gap, similar to most compounds where
superconductivity is mediated by conventional electron-phonon
interactions. However, recent experimental studies, such as
specific heat \cite{nohara,izawaYN1}, thermal conductivity
\cite{tail1},  Raman scattering \cite{raman}, and photoemission
spectroscopy \cite{yokoya}  on  YNi$_2$B$_2$C or LuNi$_2$B$_2$C
have reported a large anisotropy in the gap function. Below we
review the implications of the thermal conductivity measurements
for the gap symmetry.

\begin{figure}[t]
\begin{center}
\includegraphics[width=10cm]{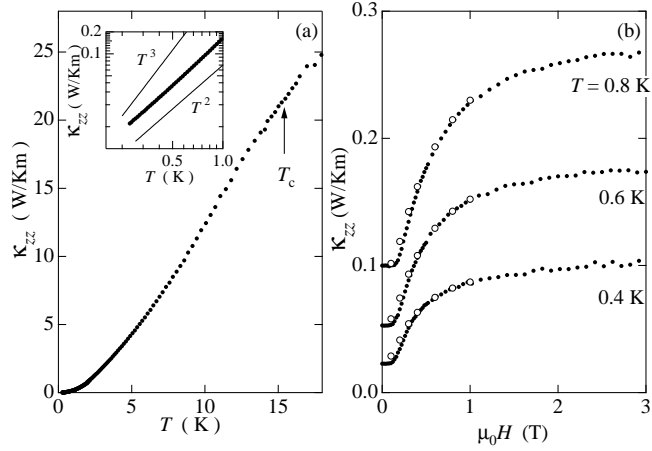}
\caption{(a)Temperature dependence of the $c$-axis thermal conductivity $\kappa_{zz}$ in zero field.  Inset: Log-log plot of the same data below 1~K.  (b) Field dependence of $\kappa_{zz}$ at low temperatures ({\boldmath $H$}$\parallel $[110]).  The solid circles represent the data measured by sweeping $H$ after zero field cooling, and the open circles represent the data measured under field cooling conditions at each temperature.}
\label{fig:YNi2B2C_T}
\end{center}
\end{figure}

Figure \ref{fig:YNi2B2C_T} (a) shows the $T$-dependence of the
$c$-axis thermal conductivity $\kappa_{zz}$ (the heat current {\bf
q} $\parallel c$) of YNi$_2$B$_2$C ($T_c$=15.5~K) single crystal
with no magnetic field.  The residual resistivity ratio of this
crystal is approximately 47 (the highest crystal quality currently
achievable).   Upon entering the superconducting state,
$\kappa_{zz}$ exhibits a small kink, as expected for a second
order transition. The Wiedemann-Franz ratio at $T_{c}$, $L =
\kappa_{zz}\rho_{zz}/T\simeq1.02L_0$ indicating that the
electronic contribution to $\kappa$ is dominant.   The inset of
Fig.\ref{fig:YNi2B2C_T} (a) shows the same data below 1~K, where
the $T$-dependence of $\kappa_{zz}$ is close to quadratic (rather
than cubic, as it would be for dominant phonon contribution).
Figure \ref{fig:YNi2B2C_T} (b) depicts the magnetic field
dependence of $\kappa_{zz}$ ({\boldmath $H$}$\parallel $[110]) at
low temperaures.  Rapid increase of $\kappa_{zz}$ at low fields is
markedly different from that observed in typical $s$-wave
materials \cite{Nb}. This steep increase of the thermal
conductivity, along with the $\sqrt{H}$-dependence of the heat
capacity \cite{nohara,izawaYN1}, strongly suggests that the
thermal properties are governed by the delocalized QPs arising
from the nodes (or extremely deep minima) in the gap.

\begin{figure}[t]
\begin{center}
\includegraphics[width=7cm]{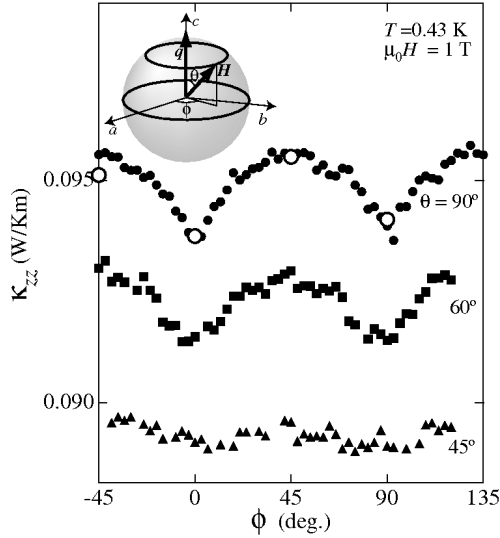}
\caption{ Angular variation of $\kappa_{zz}$ ({\bf q}$\parallel c$), measured by rotating {\boldmath
$H$}$(\theta,\phi)=H(\sin\theta\cos\phi, \sin\theta\sin\phi,\cos\theta)$ conically as a function of $\phi$ at fixed $\theta=90^{\circ}$, $60^{\circ}$, and $45^{\circ}$ (see the inset).  The open circles represent the data obtained under the field cooling condition at each angle}
\label{fig:YNi2B2C_angle}
\end{center}
\end{figure}

Having established the predominant  contribution of the extended
QPs in the thermal transport, we are in the position to address
the nodal structure of the gap function. As discussed above, in
three-dimensional systems, conical rotation of the field allows a
more direct observation of the nodal structure. Figure
\ref{fig:YNi2B2C_angle} displays the angular variation of
$\kappa_{zz}$, measured by rotating {\boldmath
$H$}=$H(\sin\theta\cos\phi, \sin\theta\sin\phi, \cos\theta)$
conically, as a function of $\phi$, at a constant $\theta$. The
measurements were done by rotating $\phi$ after field cooling at
$\phi=-45^{\circ}$. The open circles in
Fig.\ref{fig:YNi2B2C_angle} show $\kappa_{zz}(H, \phi)$ at $H$=1~T
which are obtained under the field cooling at each angle, and
demonstrate excellent agreement between the two sets of
measurements.  A clear fourfold symmetry is observed for the
$\phi$-rotation at $\theta=90^{\circ}$ and $60^{\circ}$, so that
$\kappa_{zz}=\kappa_{zz}^0+\kappa_{zz}^{4\phi}$. Here
$\kappa_{zz}^0$ is $\phi$-independent, and $\kappa_{zz}^{4\phi}$
has the fourfold symmetry with respect to $\phi$-rotation.

As seen in Fig.~\ref{fig:YNi2B2C_angle},  $\kappa_{zz}^{4\phi}$
has a narrow cusp at $\phi=0^{\circ}$ and 90$^{\circ}$. We stress
that the anisotropies of the Fermi velocity $v_F$ and $H_{c2}$,
which are inherent to the tetragonal band structure of
YNi$_2$B$_2$C, are unlikely to be at the origin of the observed
fourfold symmetry. The 4-fold $\phi$-dependence of $H_{c2}$ at
$\theta=90^{\circ}$ and 45$^{\circ}$ is nearly perfectly
sinusoidal \cite{Metlushko}, and therefore different from the $\phi$-dependence of
$\kappa_{zz}^{4\phi}$ displayed in Fig.~\ref{fig:YNi2B2C_angle}.
According to the previous section, the minima of
$\kappa_{zz}^{4\phi}$ at $\phi=0^{\circ}$ and 90$^{\circ}$
immediately indicate that {\it the nodes are located along [100]
and [010]-directions.}

The cusp structure and the $\theta$-dependence of
$\kappa_{zz}^{4\phi}$ are key features for specifying the type of
nodes. First, the cusp itself is  markedly different from the
smooth (almost sinusoidal) feature predicted (see previous
section, Fig.\ref{fig:Line_point}) and observed (see next section)
in superconductors with line nodes, such as $d$-wave. Second, the
amplitude of $\kappa_{zz}^{4\phi}$ decreases rapidly as {\boldmath
$H$} is changed from the in-plane $\theta=90^{\circ}$ to
$45^{\circ}$.
Therefore direct comparison of the data  on both the cusp
structure and $\theta$-dependence of $\kappa_{zz}$ with
Fig.\ref{fig:Line_point}) strongly favors a model with point
nodes, and leads us to conclude that the superconducting gap
function of YNi$_2$B$_2$C has {\it point nodes} along [100] and
[010]-directions.

For a gap with point nodes the $T$-dependence of the thermodynamic
quantities at low temperature depends on whether the gap increases
linearly or quadratically with the distance from the nodal point.
 {\em The gap
function we used predicts a quadratic $T$-dependence of the
thermal conductivity, which is consistent with the data in the
inset of Fig.~\ref{fig:YNi2B2C_T}(a).}

More recent measurements of the angular  variation of the heat
capacity also report the four fold oscillations consistent with
the present experiments \cite{park1}.  Very recent STS
measurements in the vortex state of YNi$_2$B$_2$C have
demonstrated  the presence of the extended QPs in the [010]
direction \cite{nishida}.  Thus the nodal structure is confirmed
by the several different techniques.

Are these real nodes or simply deep minima? Experimentally, a
clear fourfold pattern is seen at T=0.27~K and $H=1$T$\sim
0.1H_{c2}$. This suggests that the typical Doppler energy,
$E_{av}\sim\Delta_0\sqrt{H/H_{c2}}$, of the nodal quasiparticles
far exceeds both $T$, and the minimal gap $\Delta_{min}$. Here we
estimate $\Delta_0\sim \hbar v_F/\pi\xi_0$. This leads to the
anisotropy ratio $\Delta_{min}/\Delta_0\ll 0.3$. A more stringent
constraint may be deduced from the power law temperature
dependence of the thermal conductivity down to this temperature in
zero field. Estimating, $\Delta_0\sim 28$K from the value of
$T_c$, we find $\Delta_{min}/\Delta_0\lesssim 0.01$.

While this value is small, the origin of the true nodes is
topological, and hence the important question is whether the gap
function changes its sign on the Fermi surface.  To answer it, we
examined the impurity effect on the gap anisotropy. In an
anisotropic $s$-wave superconductor, with accidental gap minima or
zeroes, introduction of non-magnetic impurities affects $T_c$ only
moderately, and rapidly makes the gap more isotropic thereby
reducing the DOS at the Fermi surface by removing the node. On the
other hand, if the gap changes sign and its average over the Fermi
surface vanishes, doping with impurities suppresses $T_c$ more
severely, and induces a finite DOS at energies smaller than the
scattering rate, $\gamma$. In the latter case the oscillations of
$\kappa(H)$ persist in the regime where the Doppler energy
$\gamma\lesssim E_{av} $.

\begin{figure}[t]
\begin{center}
\includegraphics[width=8cm]{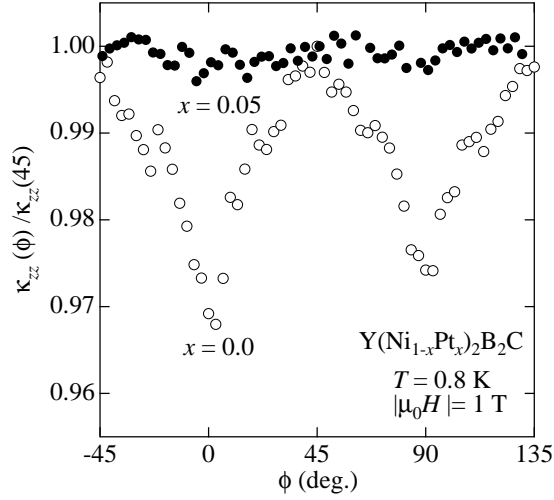}
\caption{Angular variation of $\kappa_{zz}$ for YNi$_2$B$_2$C (open circle) and Y(Ni$_{0.95}$Pt$_{0.05}$)$_2$B$_2$C (solid circles), measured by rotating {\boldmath
$H$}  as a function of $\phi$ within the $ab$-plane.  Angular variation disappears in Y(Ni$_{0.95}$Pt$_{0.05}$)$_2$B$_2$C.}
\label{fig:YNiPtBC}
\end{center}
\end{figure}

Figure \ref{fig:YNiPtBC} shows the absence of the angular
variation of the thermal conductivity
$\kappa_{zz}(\theta=90^{\circ},\phi)$ in the Pt-substituted
compound Y(Ni$_{1-x}$Pt$_x$)$_2$C with $x$=0.05. We estimate that
$E_{av}\sim 0.3\Delta_0$. On the other hand, the transition
temperature changes little with Pt-doping \cite{nohara,kamata}.
Then the disappearance of the angular variation in $\kappa$
indicates the opening of the gap, and the destruction of nodal
regions by impurity scattering. This is consistent with the heat
capacity measurements, which reports a transition from $\sqrt{H}$
behavior of $\gamma(H)$ at $x$=0 to linear in $H$ behavior for
$x$=0.2 \cite{nohara}.

It has been pointed out that the cusp structure in the angular
variation of the thermal conductivity can appear as a result of
the nesting property of the Fermi surface \cite{udagawa2}.
However,  the disappearance of the angular variation in
Y(Ni$_{1-x}$Pt$_x$)$_2$C with $x$=0.05, indicates that this
scenario is unlikely.  Moreover the cusp structure appears even in
ErNi$_2$B$_2$C, in which the nesting part of the Fermi surface
disappears due to the spin-density-wave transition \cite{kawano}.

Therefore, comparison of the experiment with existing theories
yields the gap structure with point nodes.  While these may be
accidental, an alternative view is that a strong Coulomb repulsion
is an essential ingredient of the models required for the
borocarbides. Recently, it has been shown that the $s$-wave
superconductivity with deep gap minimum appears when the electron
phonon coupling coexists with the AF fluctuation \cite{kontani}.
We note that the topology of the nodal regions plays an important
role in determining the superconducting properties, such as the
vortex lattice structure, reversible magnetization, upper critical
field $H_{c2}$, etc. For instance, the extended QPs appear to be
very important for the vortex triangular-square lattice phase
transition \cite{park2,vl,nakai3}.

\subsection{Heavy Fermion UPd$_2$Al$_3$}

UPd$_2$Al$_3$  has aroused great interest among heavy fermion (HF)
superconductors because of its unique properties.  In
UPd$_2$Al$_3$, superconductivity with heavy mass occurs at
$T_c$=2.0~K after antiferromagnetic (AF) ordering with atomic size
local moments ($\mu=$0.85$\mu_B$) sets in at $T_N$=14.3~K
\cite{geibel}.   Below $T_c$, superconductivity coexists with
magnetic ordering.  The ordered moments are coupled
ferromagnetically in the basal hexagonal $ab$-plane and line up
along the $a$-axis (Fig.9).  These ferromagnetic sheets are stacked
antiferromagnetically along the $c$-axis with the wave vector
{\boldmath $Q_0$}=(0,0,$\pi/c$), where $c$ is the $c$-axis lattice
constant \cite{kita} (For the structure of the hexagonal basal
plane, see the inset of Fig.\ref{fig:UPd2Al3_fig1}.   The presence
of large local moments is in contrast to  other HF
superconductors, in which static magnetic moments are either
absent or very small at the onset of superconductivity
\cite{thalmeier}. Since both superconductivity and AF ordering in
UPd$_2$Al$_3$ involve the Uranium 5$f$ electrons, this system is
an example of the dual nature, partly localized and partly
itinerant, of strongly correlated electrons
\cite{thalmeier,zwicknagl,sato}.

\begin{figure}[b]
\begin{center}
\includegraphics[width=6cm]{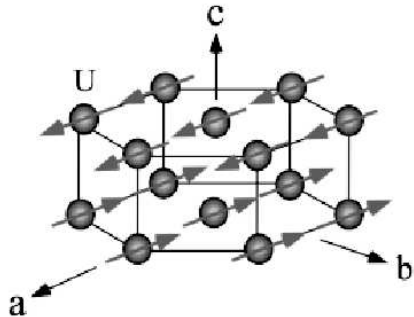}
\caption{Crystal structure of UPd$_2$Al$_3$.  AF ordering with atomic size
local moments ($\mu=$0.85$\mu_B$) sets in at $T_N$=14.3~K.  The ordered moments are coupled
ferromagnetically in the basal hexagonal $ab$-plane and line up along the $a$-axis.  These ferromagnetic sheets are stacked antiferromagnetically along the $c$-axis.}
\end{center}
\end{figure}

In the superconducting state of UPd$_2$Al$_3$,   two noticeable
features have been reported.  The first is the  ''strong coupling
anomalies" observed in the tunnel junctions
UPd$_2$Al$_3$-AlO$_x$-Pb \cite{jourdan} and inelastic neutron
scattering, which were attributed to the strong interaction
between the heavy fermion quasiparticles and AF spin-wave
excitations \cite{sato,bernhoeft1,bernhoeft2,metoki}. The second
feature is the appearance of a ''resonance peak" in the inelastic
neutron scattering in the vicinity of  {\boldmath $Q_0$} well
below $T_c$. Similar peak in the cuprates was interpreted as the
result of the feedback effect of the opening of the
superconducting gap on the electron-hole damping of the spin
fluctuations, and was shown to be unique to superconductors where
the gap changes sign under the translation by the AFM wave vector
\cite{Abanov:1999}. { The analogous effect in UPd$_2$Al$_3$, which
has static AFM order, rather than fluctuations,  was investigated
by Bernhoeft et al.\cite{bernhoeft1,bernhoeft2}, and strongly
suggests that the gap in this material changes sign under
translation {\bf k}$\rightarrow$ {\boldmath $k+Q_0$}. The NMR
Knight shift measurements indicate the spin-singlet pairing, and
the spin-lattice relaxation rate does not show the coherence peak
at $T_c$, and decreases as $T_1^{-1}\propto T^3$, indicating the
presence of line nodes \cite{touUPdAl}. These results provide
rigorous constraints on the shape of the gap.

\begin{figure}[t]
\begin{center}
\includegraphics[width=8cm]{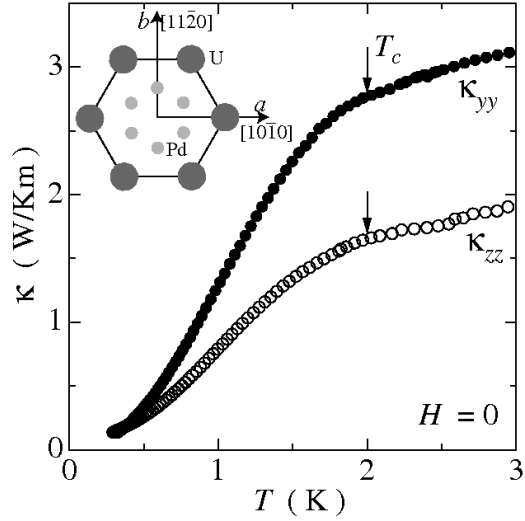}
\caption{Temperature dependence of the thermal conductivity along
the $b$-axis $\kappa_{yy}$ ({\bf q}$\parallel b$) and $c$
axis $\kappa_{zz}$ ({\bf q}$\parallel c$)in zero field.
Inset: structure of the hexagonal basal plane of UPd$_{2}$Al$_{3}$
with the alignment of the $a$-axis (100) and $b$-axis (-1,2,0), as
used in the text.  } \label{fig:UPd2Al3_fig1}
\end{center}
\end{figure}

According to band calculations and de Haas-van Alphen measurements
in the AF phase, the largest Fermi sheet with heavy electron mass
and the strongest 5f-admixture has the shape of a corrugated
cylinder with a hexagonal in-plane anisotropy \cite{zwicknagl}.
Below we assume that this cylindrical Fermi sheet with heavy mass
is responsible for the superconductivity, and carry out the
analysis within this model.

We measured the thermal conductivity  along the $c$-axis of the
hexagonal crystal structure, $\kappa_{zz}$ (heat current
{\bf q} $\parallel$ $c$) and along the $b$-axis
$\kappa_{yy}$ ({\textit {\textbf q}} $\parallel$ $b$) in high
quality single crystals of UPd$_2$Al$_3$ with $T_c=2.0$~K (The
residual resistivity ratio was 55 along the $b$-axis and 40 along
the $c$-axis).  Figure \ref{fig:UPd2Al3_fig1} depicts the
temperature dependence of $\kappa_{yy}$ and $\kappa_{zz}$ in zero
field.   Since spin-wave spectrum has a finite gap of
$\sim1.5$~meV at the zone center, its contribution appears to be
negligible below $T_c$ \cite{touUPdAl,chiao}.  The Wiedemann-Franz
ratio $L=\frac{\kappa}{T} \rho$ at $T_c$ is 0.95$L_0$ for
$\kappa_{zz}$ and is 1.16$L_{0}$ for $\kappa_{yy}$.  These results
indicate that the electron contribution is dominant below $T_c$.

\begin{figure}[t]
\begin{center}
\includegraphics[width=8cm]{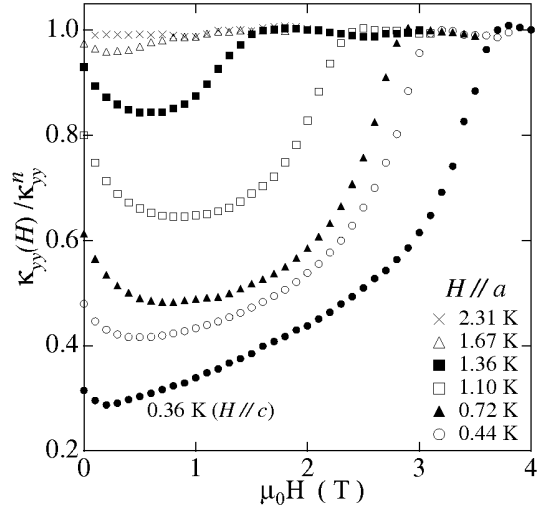}
\caption{ Field dependence of the $b$-axis thermal conductivity
$\kappa_{yy}$ normalized to the normal state value just above the
upper critical field, $\kappa_{yy}^{n}$, for {\boldmath
$H$}$\parallel a$ and {\boldmath $H$}$\parallel c$. }
\label{fig:UPd2Al3_fig2}
\end{center}
\end{figure}
Figure \ref{fig:UPd2Al3_fig2} shows the  $H$-dependence of
$\kappa_{yy}$ for {\boldmath $H$}$\parallel a$ and {\boldmath
$H$}$\parallel c$ below $T_c$.  For both field directions,
$\kappa_{yy}$ grows with $H$ beyond an initial decrease at low
fields. For  {\boldmath $H$}$\parallel c$, $\kappa_{yy}$ increases
almost linearly with $H$, $\kappa_{yy}\propto H$, at 0.36~K.  The
minimum in $\kappa(H)$ is much less pronounced at lower
temperatures. As $H$ approaches $H_{c2}$ $\parallel a$ or
$\parallel b$, $\kappa_{yy}$ shows a steep increase and attains
its normal state value.  This low to intermediate $H$-dependence
of the thermal conductivity in the superconducting state is
markedly different from that observed in ordinary $s$-wave
superconductors. At high temperatures and low fields, where the
condition $\sqrt{H/H_{c2}}<T/T_{c}$ is satisfied, the thermally
excited quasiparticles dominate over the Doppler shifted
quasiparticles. It has been shown that in this regime, while the
Doppler shift enhances the DOS, it also leads to a concomitant
reduction in both the impurity scattering time and Andreev
scattering time off the vortices
\cite{kubert,vekhter2,izawaCe,izawaSr,won}. When this lifetime
suppression exceeds the enhancement in $N(E)$, which may happen at
intermediate temperatures and low fields, the nonmonotonic field
dependence of the thermal conductivity is found.  As in other
superconductors with nodes, the region of the initial decrease of
the thermal conductivity shrinks at low $T$.   Thus the
$H$-dependence of $\kappa_{yy}$ in UPd$_2$Al$_3$,  initial
decrease at low field at high temperatures and linear behavior
$\kappa_{yy}\propto H$ at low temperatures, are in qualitative
agreement with the existence of line nodes in
$\Delta(\mbox{\boldmath $H$})$ \cite{thalUPdAl1,tewordt1}.

\begin{figure}[t]
\begin{center}
\includegraphics[width=7cm]{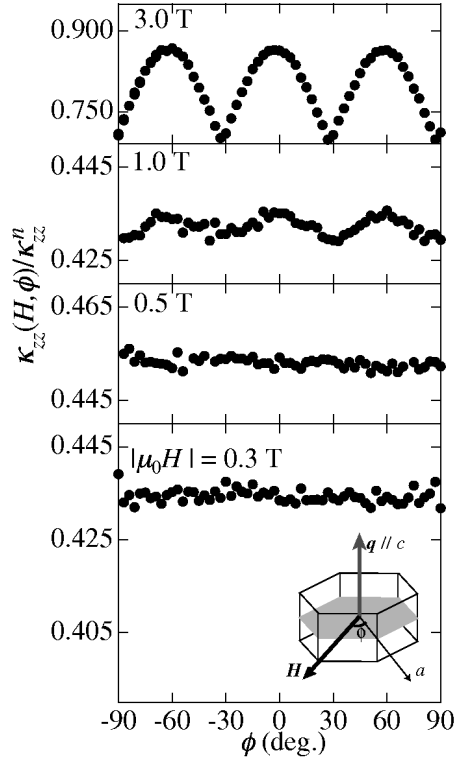}
\caption{Angular variation of the $c$-axis thermal conductivity
$\kappa_{zz}$({\boldmath $H$},$\phi$) normalized by the normal
state value, $\kappa_{zz}^n$, at several fields at 0.4~K, below
$H_{c2}$(=3.5~T).  {\boldmath $H$} was rotated within the basal
plane  as a function of $\phi$ (see the inset).  A distinct
six-fold oscillation is observed above 0.5~T, while oscillation is
absent at 0.5 and 0.3~T.  } \label{fig:UPd2Al3_fig3}
\end{center}
\end{figure}

We first test whether there exist vertical line nodes
perpendicular to the basal plane.   Figure \ref{fig:UPd2Al3_fig3}
shows $\kappa_{zz}$({\boldmath $H$},$\phi$) as a function of
$\phi$ at 0.4~K,  measured by rotating {\boldmath $H$} within the
basal plane $(\theta=90^{\circ})$.   Above 0.5~T, a distinct
six-fold oscillation is observed in $\kappa_{zz}$({\boldmath
$H$},$\phi$), reflecting the hexagonal symmetry of the crystal.
Six-fold oscillation is observable even above $H_{c2}$.  On the
other hand,  no discernible six-fold oscillation was observed
below 0.5~T within our experimental resolution.  We infer that the
AF magnetic domain structure and anisotropy of $H_{c2}$ within the
plane are responsible for the six-fold symmetry and the nodal
structure is not related to the oscillation.  According to the
neutron diffraction experiments, the magnetic domain structure
changes at $H_D\sim$0.6~T well below $T_N$.  Below $H_D$, the
ordered moments point to the $a$-axis, forming domains, and the
spin structure is not affected by the {\boldmath $H$}-rotation in
the basal plane.    On the other hand, above $H_D$, the {\boldmath
$H$}-rotation causes domain reorientation.  Then the magnetic
domain structure changes with sixfold symmetry with {\boldmath
$H$}-rotation.  Thus the sixfold symmetry observed in
$\kappa_{zz}$ above 0.5~T is most likely to be due to the magnetic
domain structure.   This indicates that {\it there are no nodes
located perpendicular to the basal plane, i.e. gap function in the
basal plane is isotropic.}

\begin{figure}[t]
\begin{center}
\includegraphics[width=8cm]{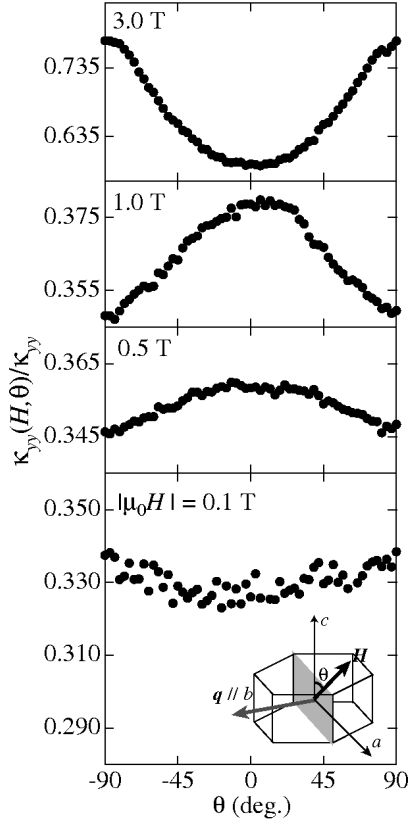}
\caption{Angular variation of the $b$-axis thermal conductivity
$\kappa_{yy}$({\boldmath $H$},$\theta$) normalized by the normal
state value $\kappa_{yy}^n$ at several fields at 0.4~K.
{\boldmath $H$} was rotated within the $ac$-plane perpendicular to
the basal plane (see the inset).} \label{fig:UPd2Al3_fig4}
\end{center}
\end{figure}

We next test for the existence of the horizontal line nodes
parallel to the basal plane.  Figure \ref{fig:UPd2Al3_fig4}
displays the angular variation of $\kappa_{yy}$({\boldmath
$H$},$\theta$) for rotating {\boldmath $H$} as a function of
$\theta$ within the $ac$-plane at 0.4~K.  A distinct oscillation
with two-fold symmetry was observed in the superconducting state.
In contrast to $\kappa_{zz}$, no discernible twofold oscillation
was found in the normal state above $H_{c2}$.   We decompose
$\kappa_{yy}$({\boldmath $H$},$\theta$) as
\begin{equation}
\kappa_{yy}=\kappa_{yy}^0+\kappa_{yy}^{2\theta},
\end{equation}
where $\kappa_{yy}^0$ is a $\theta$-independent term and
$\kappa_{yy}^{2\theta}=C_{yy}^{2\theta}\cos{2\theta}$ is a term
with the twofold symmetry with respect to $\theta$-rotation. The
field dependence of $C_{yy}^{2\theta}$ at 0.4~K is shown in Figure
\ref{fig:UPd2Al3_fig5} (a).  For comparison, the $H$-dependence of
$\kappa_{yy}$ for  {\boldmath $H$}$\parallel c$ at $T$=0.36~K is
plotted in Fig.\ref{fig:UPd2Al3_fig5} (b). There are three regions
denoted (I), (II) and (III), below $H_{c2}$. In the vicinity of
$H_{c2}$ ((III)-region), where $\kappa_{yy}$ increases steeply
with $H$, the sign of $C_{yy}^{2\theta}$ is negative and the
amplitude $|C_{yy}^{2\theta}|/\kappa_{yy}^n$ is of the order of
10\%.  Here $\kappa_{yy}^n$ is  $\kappa_{yy}$ in the normal state
just above $H_{c2}$.   With decreasing $H$, $C_{yy}^{2\theta}$
changes  sign at about 2.3~T and becomes positive in the region
where $\kappa_{yy}$ for {\boldmath $H$}$\parallel c$ shows a
linear $H$-dependence ((II)-region). Below about 0.25~T, where the
second sign change takes place, $\kappa_{yy}$ decreases with $H$
((I)-region).  In this region, $C_{yy}^{2\theta}$ is, once again,
negative.

We address the origin of the observed two-fold oscillation. The
disappearance of the oscillation above $H_{c2}$, together with the
fact that there is only one magnetic phase in this configuration
\cite{kita}, completely rule out the possibility that the origin
is due to the magnetic domain structure.   There are two possible
origins for the oscillation; the nodal structure and the
anisotropy of the Fermi velocity and $H_{c2}$.  Obviously, as
discussed previously, a large two-fold oscillation with negative
sign observed in the (III)-region arises from the anisotropies of
the Fermi velocity and  $H_{c2}$.   This immediately indicates
that {\it the two-fold symmetry with positive sign in the
(II)-region originates not from these anisotropies but from the
quasiparticle structure associated with the nodal gap function.}
In addition, the amplitude of $C_{yy}^{2\theta}/\kappa_{yy}$ in
the (II)-region is a few percent, which is quantitatively
consistent with the prediction based on the Doppler shifted DOS.
We also note that the second sign change at low fields in the
(I)-region is compatible with the nodal structure.  In this
region, as discussed previously, the $H$-dependence of the thermal
conductivity is governed by the suppression of the quasiparticle
scattering rate.

As discussed in Ref. \cite{vekhter3,yu,aubin,izawaCe,ocana,won},
the anisotropic carrier scattering time associated with the nodal
structure also gives rise to the variation of
$\kappa_{yy}$({\boldmath $H$},$\theta$) as a function of $\theta$.
In this case the sign of the oscillation is opposite to that
arising from the Doppler shifted DOS in the (II)-region.   These
considerations lead us to conclude that UPd$_2$Al$_3$ has
horizontal nodes.  In addition, the fact that there is a single
maximum structure in the angular variation of
$\kappa_{yy}$({\boldmath $H$},$\theta$) indicates that horizontal
line nodes are located at  positions where the condition {\bf
p}$\parallel ab$ in the  Brillouin zone is satisfied. Thus {\it
the allowed positions of the horizontal nodes are restricted at
the bottleneck and  AF zone boundary} \cite{watanabe,thalUPdAl1}.

\begin{figure}[t]
\begin{center}
\includegraphics[width=8cm]{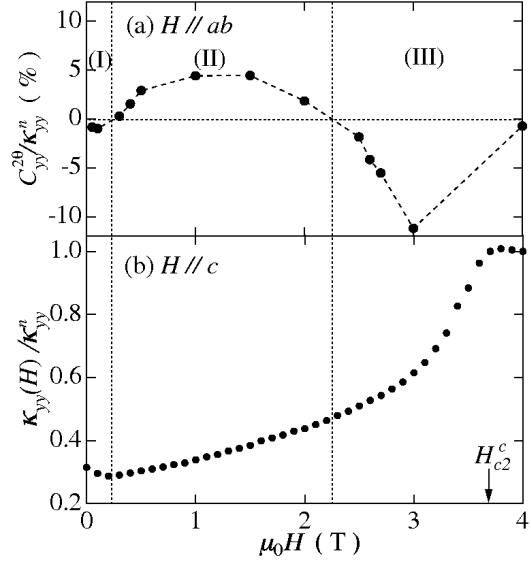}
\caption{(a) Field dependence of the amplitude of the two-fold
symmetry  $C_{yy}^{2\theta}$ normalized by the normal state
thermal conductivity $\kappa_{yy}^n$ at 0.4~K.  (b) Field
dependence of the $b$-axis thermal conductivity $\kappa_{yy}$ for
{\boldmath $H$}$\parallel c$ at 0.36~K.   The sign of
$C_{yy}^{2\theta}$ is negative in the (III)-region just below
$H_{c2}$ and in the (I)-region where $\kappa_{yy}$ decreases with
$H$.  On the other hand,  the sign of $C_{yy}^{2\theta}$ is
positive in the (II)-region. } \label{fig:UPd2Al3_fig5}
\end{center}
\end{figure}

\begin{figure}[t]
\begin{center}
\includegraphics[width=8cm]{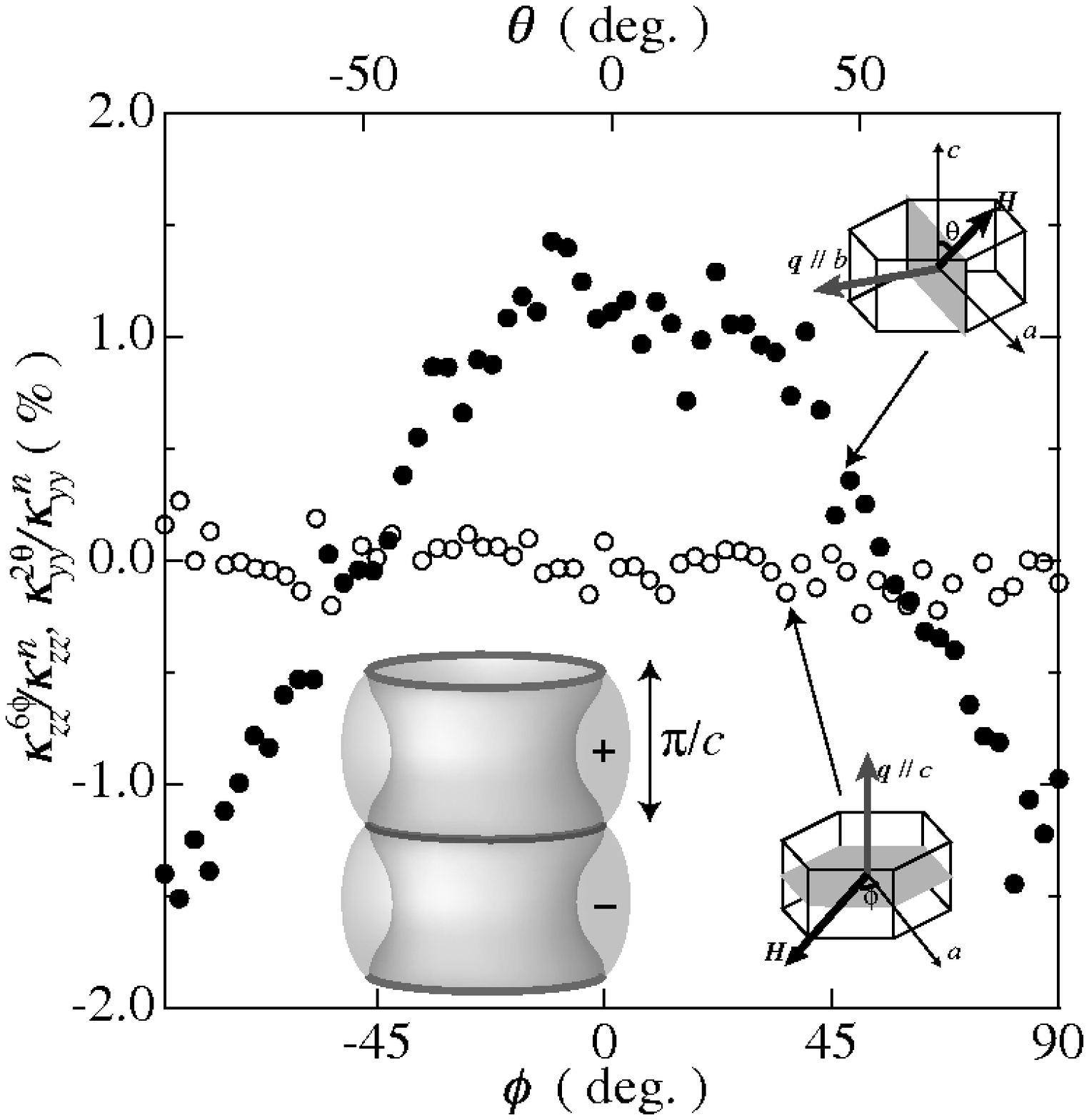}
\caption{Angular variation of the $b$-axis thermal conductivity
$\kappa_{yy}^{2\theta}/\kappa_{yy}^{n}$ with rotating {\boldmath
$H$} within the $ac$-plane (solid circles) and of the $c$-axis
thermal conductivity $\kappa_{zz}^{6\phi}/\kappa_{zz}^{n}$ with
rotating {\boldmath $H$} within the basal $ab$-plane (open
circles) at $T$=0.4~K and at $H$=0.5~T.  The amplitude of the
six-fold oscillation in  $\kappa_{zz}^{6\phi}/\kappa_{zz}^{n}$ is
less than 0.2\% if it exists.  Inset: Schematic figure of the gap
function of UPd$_{2}$Al$_{3}$ determined by angle resolved
magnetothermal transport measurements. The thick solid lines
indicate horizontal nodes located at the AF zone boundaries.   }
\label{fig:UPd2Al3_fig6}
\end{center}
\end{figure}
For comparison,  the angular variations  of $\kappa_{yy}$ and
$\kappa_{zz}$ at low fields are shown in
Fig.\ref{fig:UPd2Al3_fig6}.    While the amplitude of the two-fold
oscillation $C_{yy}^{2\theta}/\kappa_{yy}$ is 3\%, which is
quantitatively consistent with the Doppler shifted DOS, the
amplitude of the six-fold oscillation
$C_{zz}^{6\theta}/\kappa_{zz}$ is less than 0.2\%, which is more
than 10 times smaller than the amplitude expected from the Doppler
shifted DOS in the presence of nodes.    Combining the results, we
arrive at the conclusion that {\it the gap function is isotropic
in the basal plane and has horizontal node.}

To discuss the position of the horizontal line node in
UPd$2$Al$_3$,  we consider a ''magnetic"  Brillouin zone in a
cylindrical Fermi surface, as shown in
Fig.\ref{fig:Doppler_shift_h} (a).  The density of states is
anisotropic under the rotation of the field in the $ac$-plane, by
varying the angle $\theta$ in the inset.  To illustrate the
difference between the line nodes at high symmetry positions in
the magnetic Brillouin zone, and away from those, we consider here
four model gap functions
\begin{enumerate}
\item type-I : A horizontal node located at the bottleneck;
$\Delta(\mbox{\bf k}) \propto \sin k_zc$.
\item type-II : A horizontal node located at the zone boundary;
$\Delta(\mbox{\bf k}) \propto \cos k_zc$.
\item type-III: A hybrid of type-I and -II. Two horizontal nodes
located  at the bottleneck and the zone boundary;
$\Delta(\mbox{\bf k}) \propto \sin 2k_zc$.
\item type-IV : Two horizontal nodes located at
positions shifted off the bottleneck in the Brillouin zone;
$\Delta(\mbox{\bf k}) \propto \cos 2k_zc$.
\end{enumerate}
The expected angular variation of the Doppler shifted DOS is a
function of the relative angle between {\boldmath $H$} and
{\bf p} for these gap functions are shown schematically in
Fig.\ref{fig:Doppler_shift_h} (b).   The twofold oscillations with
the same phase are expected for type-I, -II , and -III gap
functions, in which the horizontal nodes are located at the
position where {\bf p}$\parallel ab$-plane; one cannot
distinguish these three gap functions when the Fermi surface has
an open orbit along the $c$-axis.  For type-IV, one expects an
oscillation with a double minimum structure as a function of
$\theta$.

\begin{figure}[t]
\begin{center}
\includegraphics[width=14cm]{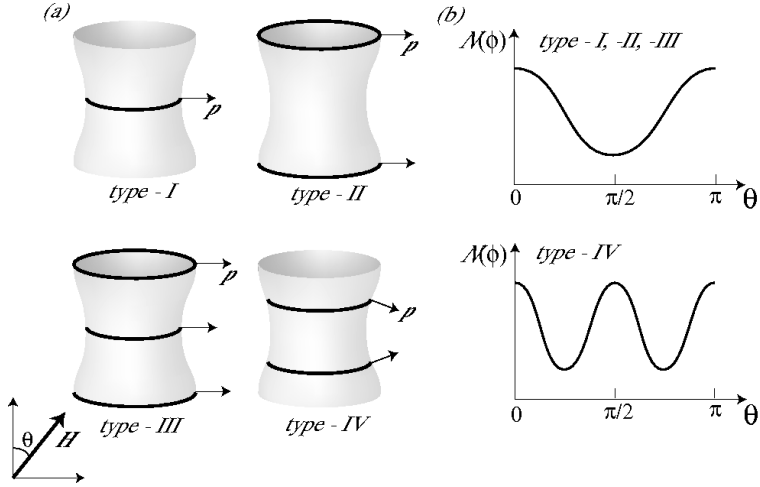}
\caption{ (a) Schematic figure of the gap structure with horizontal line node in the magnetic Brillouin zone.  Line nodes are located at the bottleneck (type-I) and at the zone boundary (type-II).  Two line nodes are located at the bottleneck and the zone boundary (type-III) and at positions shifted off the bottleneck (type-IV) .   (b) Oscillations of the DOS for {\boldmath $H$} rotating in the $ac$-plane for various gap functions.  Two-fold oscillation with the same sign are expected for type-I, -II, and -III.  On the other hand, for type-IV, an oscillation with a double minimum is expected.}
\label{fig:Doppler_shift_h}
\end{center}
\end{figure}

Thus, the order
parameters allowed,  by  thermal conductivity measurements,  are,
\begin{enumerate}
\renewcommand{\labelenumiii}{(\roman{enumi})}
\item $\Delta(\mbox{\bf k})=\Delta_0 \sin k_zc$,
\item $\Delta(\mbox{\bf k})=\Delta_0 \sin 2k_zc$   ~~~and
\item $\Delta(\mbox{\bf k})=\Delta_0 \cos k_zc$.
\end{enumerate}
which are shown in the type-I, -II and -III gap  structures in
Fig.\ref{fig:Doppler_shift_h}(a).   Generally the first and the
second represent spin triplet gap functions, and only the third,
which is a spin singlet, remains a viable possibility. Note,
however, that the thermal conductivity is only sensitive to the
{\em amplitude} of the gap. Among the possible order parameters
for the $D_{6h}$ symmetry group is that transforming according to
$\Gamma_5$ representation, with a basis function $k_z(k_x+i k_y)$
\cite{sigrist}. In that case the gap function may vary as
$(k_x+ik_y)\sin k_zc $, and the gap amplitude, over a quasi-two
dimensional Fermi surface, may have only weak modulations apart
from the horizontal line of nodes. Such an order parameter breaks
the time reversal symmetry, and we are not aware of any evidence
in support of that in UPd$_2$Al$_3$; however, targeted search for
a time-reversal symmetry broken state in this system has not been
performed. Both the first and the third gap functions are
compatible with the constraint implied by the neutron resonance
peak, $\Delta$({\bf k}) $= -\Delta$({\bf k+Q$_0$}).
These considerations lead us to conclude that {\it the gap
function of UPd$_2$Al$_3$ is most likely to be
$\Delta(\mbox{\bf k})=\Delta_0 \cos k_zc$}, shown in the
inset of Fig.\ref{fig:UPd2Al3_fig6}, although we cannot exclude
the possibility of the state with broken time-reveral symmetry on
the basis of out measurements.

For the $\cos k_zc$ pairing, the horizontal node located at the AF
zone boundary indicates that pair partners cannot reside in the
same basal plane.   The interlayer pairing appears to indicate
that strong dispersion of the magnetic excitation along $k_{z}$
causes the pairing, as suggested in the magnetic exciton mediated
superconductivity model \cite{thalmeier,mchale,sato}. The
isotropic gap function in the basal plane implies that  the
pairing interaction in the neighboring planes strongly dominates
over the interaction in the same plane.    Although the pairing
interaction inferred from the determined gap function should be
further scrutinized,  the recent results imply that the interlayer
pairing interaction associated with the  AF interaction is most
likely to be the origin of the unconventional superconductivity in
UPd$_2$Al$_3$.

\subsection{skutterudite PrOs$_4$Sb$_{12}$}

\begin{figure}[b]
\begin{center}
\includegraphics[width=6cm]{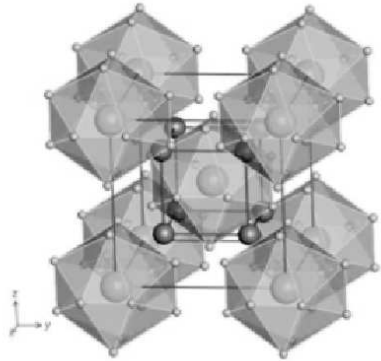}
\caption{Crystal structure of PrOs$_4$Sb$_{14}$}
\end{center}
\end{figure}

Recently discovered heavy fermion superconductor PrOs$_4$Sb$_{12}$
($T_c$=1.82~K) with filled skutterudite structure (Fig.17) is relatively
unique as the $f$-electrons have a non-magnetic ground state,
determined by the crystalline electric field (most likely singlet
$\Gamma_1$ state) \cite{bauer,aoki4}.  The HF behavior
($m*\sim50m_e$, $m_e$ is the free electron mass) is likely due to
the interaction of the electric quadrupole moments of Pr$^{3+}$
(rather than local magnetic moments as in the other HF
superconductors) with the conduction electrons.  Therefore the
relation between the superconductivity and the orbital
(quadrupole) fluctuations of $f$-electron state  excited great
interest: PrOs$_4$Sb$_{12}$ has been proposed as a candidate for
the first superconductor with pairing mediated neither by
electron-phonon nor magnetic interactions, but by quadrupolar
fluctuations. Even if these fluctuations do not provide the
pairing glue by themselves, but only in conjunction with phonons,
they have the potential for influencing the symmetry of the
superconducting order parameter, which makes it of the utmost
importance to determine the symmetry of the SC gap.

The unconventional superconductivity in  PrOs$_4$Sb$_{12}$ has
been suggested by several experiments.  NQR measurements  showed
the absence of Hebel-Slichter peak \cite{kotegawa}.  In addition,
the Knight shift does not change below $T_c$, implying that the
gap function has odd parity\cite{tou}.  Moreover, $\mu$SR
experiments report the appearance of the static spontaneous
magnetic field below $T_c$, which can be interpreted as the
spontaneous breaking of the time reversal symmetry \cite{mSRPr}.
The penetration depth and NMR $T_1^{-1}$ measurements indicate the
presence of point nodes \cite{chia}.

\begin{figure}[t]
\begin{center}
\includegraphics[width=8cm]{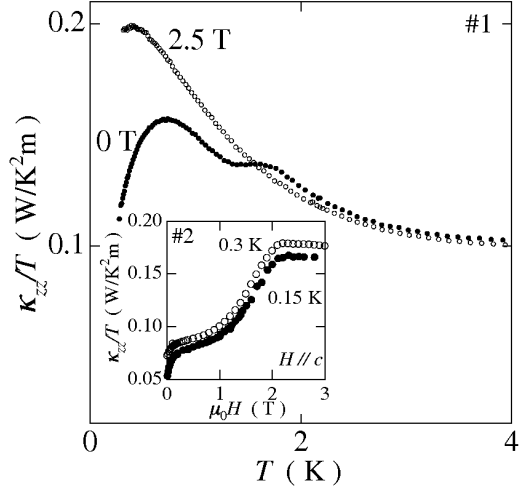}
\caption{Temperature dependence of the $c$-axis thermal conductivity $\kappa_{zz}$ (the heat current
{\bf q} $\parallel c$) divided by $T$  at zero field (solid circles) and at 2.5~T (open circles)  above $H_{c2}$ ($\simeq$ 2.2~T at $T$= 0~K) of PrOs$_4$Sb$_{12}$ single crystal ($\sharp$1).   Inset: Field dependence of $\kappa_{zz}/T$ of sample $\sharp$2 at very low temperature.
}
\label{fig:PrOs4Sb12_fig1}
\end{center}
\end{figure}

Figure \ref{fig:PrOs4Sb12_fig1} shows the  $T$-dependence of the
$c$-axis thermal conductivity $\kappa_{zz}$ (the heat current
{\bf q} $\parallel c$) divided by $T$ both at zero field
and above $H_{c2}$ ($\simeq$ 2.2~T at $T$= 0~K) of
PrOs$_4$Sb$_{12}$ single crystal ($\sharp$1).   In this
temperature region, the electronic contribution to $\kappa_{zz}$
dominates  the phonon contribution.  The inset of
Fig.\ref{fig:PrOs4Sb12_fig1}  shows the field dependence of
$\kappa_{zz}$ of sample $\sharp$2 at very low temperature.
$\kappa_{zz}$ increases very steeply even at very low field
$(H<0.1$T). When contrasted with the exponentially slow increase
of the thermal conductivity with field observed in $s$-wave
superconductors at  $H\ll H_{c2}$ \cite{brison},  this is a strong
indication that the thermal transport is governed by the
delocalized QPs arising from the gap nodes.  Above 0.1~T
$\kappa_{zz}$ increases gradually, then shows a steep increase
above 0.5~T up to $H_{c2}$.  

\begin{figure}[t]
\begin{center}
\includegraphics[width=10.5cm]{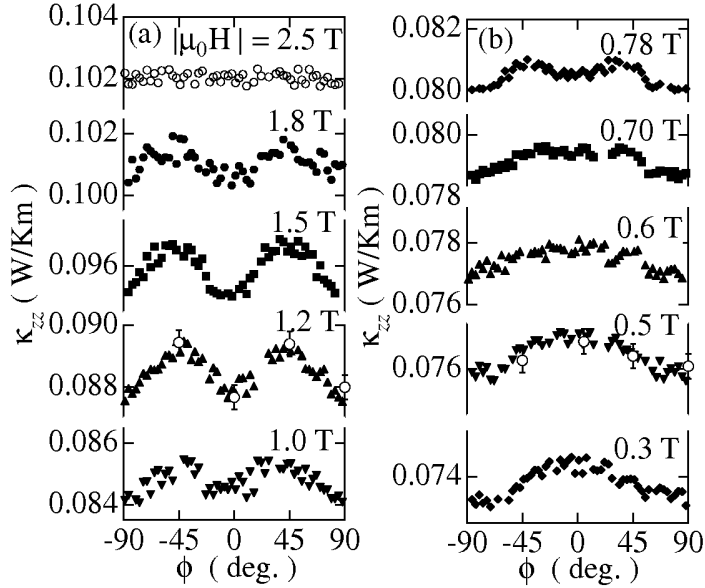}
\caption{(a)(b) Angular variation of $\kappa_{zz}$ ({\bf q}$\parallel c$) in  {\boldmath $H$} rotating within the $ab$-plane as a function of $\phi$ at 0.52~K above and below $H_{c2}(\simeq2.0$~T).  }
\label{fig:PrOs4Sb12_fig2}
\end{center}
\end{figure}

Figures \ref{fig:PrOs4Sb12_fig2} (a) and (b)  display the angular
variation of $\kappa_{zz}(${\boldmath $H$}$,\phi)$ in  {\boldmath
$H$} rotated within the $ab$-plane ($\theta=90^{\circ}$) at
$T$=0.52~K \cite{izawaPr}.  The measurements have been done in
rotating  {\boldmath $H$} after field cooling at
$\phi=-90^{\circ}$.  The open circles show
$\kappa_{zz}(${\boldmath $H$}$,\phi)$ at $H$=1.2~T and 0.5~T which
are obtained under field cooling at each angle. Above $H_{c2}$
($\simeq$2.0~T at 0.5~K) $\kappa_{zz}(${\boldmath $H$}$,\phi)$ is
essentially independent of $\phi$.  A clear fourfold variation is
observed just below $H_{c2}$ down to $H \sim$ 0.8~T.  However
further reduction of $H$ below 0.8~T causes a rapid decrease of
the amplitude of the fourfold term, and its disappearance below
0.7~T. At the same time, the twofold component grows rapidly. This
surprising behavior suggests a change in the gap symmetry as a
function of field in the superconducting state.

Figure \ref{fig:PrOs4Sb12_fig3} shows the $H$-dependence  of the
amplitudes of the twofold and the fourfold terms, which are
obtained by decomposing $\kappa_{zz}(${\boldmath $H$}$,\phi)$ as
\begin{equation}
\kappa_{zz}({\boldmath H},\phi)=\kappa^0_{zz}+\kappa^{2\phi}_{zz}+\kappa^{4\phi}_{zz},
\end{equation}
where $\kappa^0_{zz}$ is a $\phi$-independent term,
$\kappa^{2\phi}_{zz}=C_{2\phi}\cos2\phi$, and
$\kappa^{4\phi}_{zz}=C_{4\phi}\cos4\phi$ are the terms with
twofold and fourfold symmetry with respect to $\phi$-rotation. It
is clear that the transition from the fourfold to twofold symmetry
in $\phi$-rotation is sharp, and  occurs in a narrow field range
at $H/H_{c2}\simeq$ 0.4, deep inside the SC phase. Both symmetries
coexist in a narrow field range.  If the minima of the thermal
conductivity are associated with the direction of the field along
the nodes, the reduced $\kappa_{zz}(${\boldmath $H$}$,\phi)$ at
$\phi=\pm90^{\circ}$ and 0$^{\circ}$ in the  high field phase, and
at $\phi=\pm90^{\circ}$ for the low field phase, respectively,
lead to the conclusion that the nodes are along the [100]- and
[010]-directions in the high field phase, while they are located
only along the [010]-direction in the low field phase.

\begin{figure}[t]
\begin{center}
\includegraphics[width=9cm]{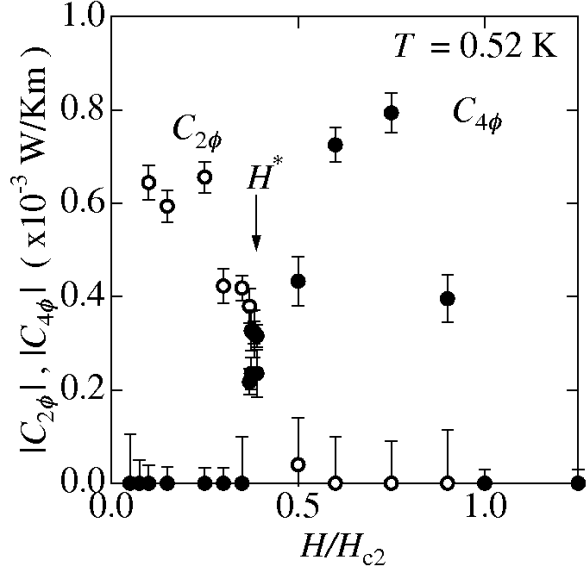}
\caption{The amplitude of twofold (open circles) and fourfold (filled circles) symmetries, $|C_{2\phi}|$ and $|C_{4\phi}|$, respectively,  plotted as a function of $H/H_{c2}$ at $T$=0.52~K. At $H^{\ast}$, the crossover from twofold to fourfold symmetry takes place.  }
\label{fig:PrOs4Sb12_fig3}
\end{center}
\end{figure}

Having established the presence of nodes, the  next question is
their classification.  As discussed in $\S$ IV A, the angular
variation of $\kappa_{zz}$ can distinguish between the point and
line nodes, by rotating  {\boldmath $H$}  conically around the
$c$-axis with a tilted angle from the $ab$-plane.  Although we do
not show it here, the amplitude at $\theta=45^{\circ}$ and
30$^{\circ}$ is smaller than that at $\theta=90^{\circ}$
\cite{izawaPr}.   Similar results were obtained for the twofold
symmetry.

What is the nodal structure inferred from the  present results?
The $\phi$-rotation of the field can only provide the information
of the nodes away from the [001] direction. We shall therefore
appeal to the group theoretical consideration for the discussion
of the nodal structure. It is unlikely that the SC gap function
has only four point nodes in the cubic $T_h$ crystal symmetry:
this is independent of the spin singlet or triplet symmetry. Hence
the likely scenario is that the gap function at high field phase
has six point nodes. The low field phase is likely to have two
nodes, although on the basis of the experimental observations we
cannot exclude the 4-node structure (along [001] and [010])

The $H-T$ phase diagram of the SC symmetry determined by the
present experiments is displayed in Fig.~\ref{fig:PrOs4Sb12_fig4}.
The filled circles represent the magnetic field $H^*$ at which the
transition from fourfold to twofold symmetry takes place. The
$H^{*}$-line which separates two SC phases (high field $A$-phase
and low field $B$-phase) lies deep inside the SC state. We note
that recent flux flow resistivity measurements also reported an
anomaly at $H^*$.  The only example of a superconductor with
multiple phases of different gap symmetry so far has been UPt$_3$
\cite{upt3}. In that case the degenerate transition temperatures
for the two orders at zero field can be split by, for example,
applying pressure. Therefore it is important to determine a)
whether the two phases manifested in the thermal conductivity
measurements have the same $T_c$ in zero field; b) whether the
transition can be split by influencing the system by an
experimental handle other than the field. It seems logical that,
if the gap structure suggested here is indeed realized,
application of the uniaxial pressure along the [100] direction
should lift the symmetry of the gap and favor one of the two
phases.

\begin{figure}[t]
\begin{center}
\includegraphics[width=8cm]{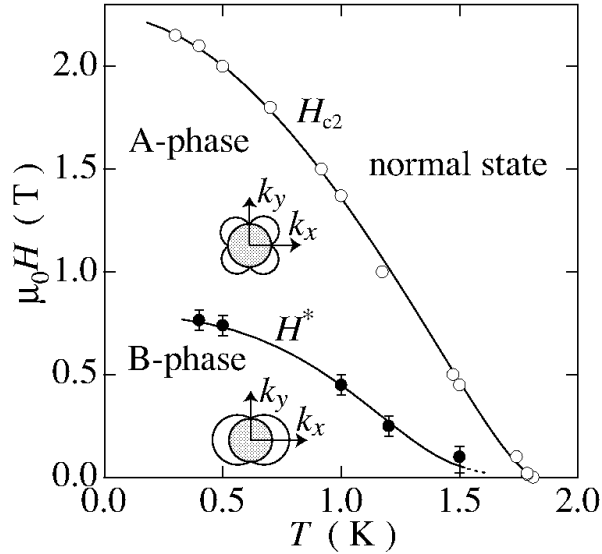}
\caption{The phase diagram of the superconducting gap symmetry determined by the present experiments.  The filled circles represent the magnetic field $H^*$ at which the transition from fourfold to twofold symmetry takes place.   The open circles represent $H_{c2}$.  The area of the gap function with fourfold symmetry is shown as $A$-phase and the area of the gap function with twofold symmetry is shown as $B$-phase. }
\label{fig:PrOs4Sb12_fig4}
\end{center}
\end{figure}

Recently, small angle neutron scattering experiments  reported the
hexagonal flux line lattice, which is distorted with a twofold
symmetry \cite{huxley}.  It has been pointed out that the
distortion originates from the nodal gap structure, which provides
a strong support of the present angular variation of
$\kappa_{zz}(${\boldmath $H$}$,\phi)$.  We also note that the
possible existence of the third phase was predicted by the
magnetization and penetration depth measurements \cite{chia,mota}.

Thus PrOs$_4$Sb$_{12}$ has several unique features.   In almost
all superconducting (SC) materials  known to date, once the energy
gap in the spectrum of electrons opens at the SC transition, only
its overall amplitude, and not the shape and symmetry around the
Fermi surface, changes in the SC phase \cite{pok}.  In contrast,
PrOs$_4$Sb$_{12}$  seems to have several superconducting phases
with different symmetries.  Many heavy fermion superconductors
have {\it line} nodes in the gap functions (with possible
additional point nodes).  The suggestion that PrOs$_4$Sb$_{12}$ is
the first heavy fermion superconductor, in which only the point
nodes are identified may be the key for understanding the
superconductivity mechanism due to quadrupolar interaction.

\section{Quasi two dimensional superconductors}

We now discuss the implications of the thermal conductivity
measurements for the nodal structure of three superconductors,
CeCoIn$_5$, $\kappa$-(BEDT-TTF)$_2$Cu(NCS)$_2$, and Sr$_2$RuO$_4$.
These materials look very different at first sight but reveal
several similar features.  First, all three have strong
electron-electron correlations. Second, they all have quasi-two
dimensional electronic structure, as confirmed by the band
structure calculation and by the dHvA measurements. This is also
supported by a relatively large anisotropy of the upper critical
field between inequivalent crystalline directions. Third, the
power laws in the  temperature dependence of thermodynamic
quantities in the superconducting state is consistent with the
presence of line nodes in the superconducting gap.  The position
of the line nodes is still an open question in many of these
materials, and is the focus of our analysis here.

Unfortunately, in all  these compounds,  the out-of-plane thermal
conductivity is difficult to measure due to thin plate-like
single-crystal samples.  We therefore measured the in-plane
thermal conductivity with magnetic field rotated within the same
conducting plane.  In this geometry,  the dominant signal is
twofold symmetric, and  simply depends on  the angle between the
thermal current, {\bf q}, and {\bf H} due to the difference in the
transport along and normal to the vortices. This twofold
oscillation is not directly related to the nodal structure, and
the challenge for the interpretation is to separate it from the
features due to the nodes.

\subsection{CeCoIn$_5$}

\begin{figure}[b]
\begin{center}
\includegraphics[width=5cm]{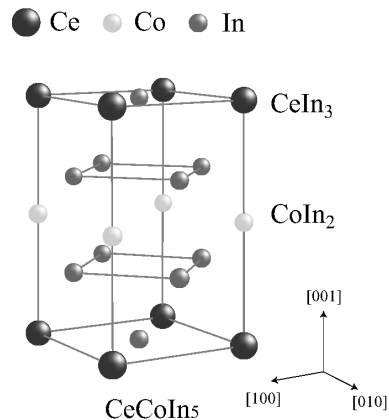}
\caption{Crystal structure of CeCoIn$_5$.}
\end{center}
\end{figure}

The family of the heavy fermion superconductors  Ce$T$In$_5$
($T$=Rh, Ir, and Co) was discovered in 2001 \cite{pet}.  Both
CeIrIn$_5$ and CeCoIn$_5$ are ambient pressure superconductors,
with transition temperatures of 0.4~K and 2.3~K, respectively.
CeRhIn$_5$ is an antiferromagnet, but shows superconductivity
under moderate pressure. The crystal structure of CeTIn$_5$ is
tetragonal consisting of the conducting CeIn$_3$ layers separated
by less conducting $T$In$_2$ layers (Fig.~22).   In the normal
state CeCoIn$_5$ exhibits non-Fermi-liquid behavior, likely
related to the strong AFM fluctuations; moreover, there is
evidence that superconductivity appears in the neighborhood of a
quantum critical point (QCP), possibly of AFM origin
\cite{sidorov,curro,bianch1,nakajima}. For that reason CeCoIn$_5$
is an excellent candidate to study the relationship between QCP
and unconventional superconductivity.  The temperature dependence
of the heat capacity and thermal conductivity \cite{mov}, thermal
Hall conductivity \cite{Kasahara}, NMR relaxation rate
\cite{kohoriCo} all indicate the presences of line nodes. The
penetration depth measurements \cite{ozcan,oremeno,chia2} in
general support this conclusion although the low-$T$ exponents are
anomalous. Recent Andreev reflection measurements indicate the
sign change of the superconducting order parameter
\cite{WKPark1,rourke,WKPark2}, see, however \cite{GSheet} for
comments on Ref.~\cite{rourke}. The Knight shift measurement of
CeCoIn$_5$ indicates the even-spin parity \cite{kohoriCo}.

\begin{figure}[t]
\begin{center}
\includegraphics[width=8cm]{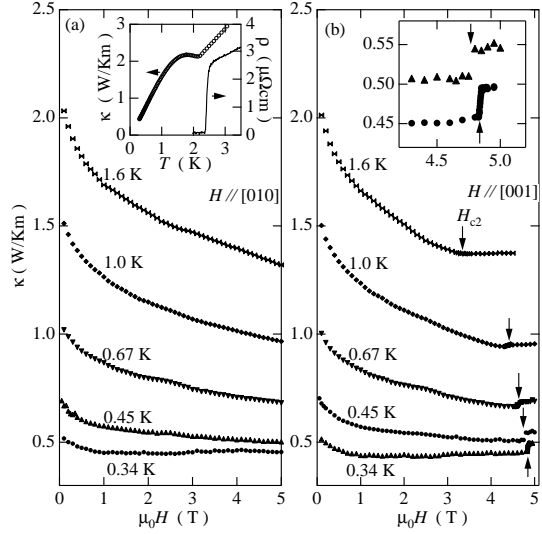}
\caption{Field dependence of the $a$-axis thermal conductivity ({\bf q}$\parallel $[100]) for (a) {\boldmath $H$}$\parallel [001]$ and (b) {\boldmath $H$}$\parallel [010]$ below $T_c$.   Inset of (a) : $\kappa$ and $\rho$ in zero field.  Inset of (b) : Field dependence of $\kappa$ near $H_{c2}$ at 0.45~K ($\circ$) and 0.34~K ($\bullet$).   The thermal conductivity jumps at $H_{c2}$, indicating the first order phase transition.}
\label{fig:CeCoIn5_fig1}
\end{center}
\end{figure}

There are indications that the upper critical field in CeCoIn$_5$
is paramagetically (Pauli) limited. At low $T$ the phase
transition at $H_{c2}$ is first order, as revealed by a step in
the $H$-dependence of the thermal conductivity, magnetization, and
specific heat \cite{izawaCe,tayama,bianch2}.  Moreover,
measurements of heat capacity\cite{FFLO1,FFLO2},
ultrasound\cite{watanabe2}, and NMR \cite{kakuyanagi1}revealed a
new superconducting phase at low temperatures in the vicinity of
$H_{c2}$ at low temperatures (${\boldmath H} \parallel ab$). This
new phase was conjectured to be the spatially inhomogeneous
superconducting state (Fulde-Ferrell-Larkin-Ovchinnikov state),
which was predicted 4 decades ago, not previously observed.

The inset of Fig.\ref{fig:CeCoIn5_fig1}(a) shows  the
$T$-dependence of $\kappa$ and $\rho$.   Upon entering the
superconducting state, $\kappa$ exhibits a sharp kink and rises to
the maximum value at $T \sim$ 1.7~K. The Wiedemann-Franz ratio $L
= \frac{\kappa}{T}\rho\simeq1.02L_0$ at $T_{c}$ is very close to
the Lorenz number $L_{0}= 2.44\times 10^{-8}$~$\Omega$W/K,
indicating that the electronic contribution is dominant. Therefore
the enhancement of $\kappa$ below $T_{c}$ is due to the
suppression of the inelastic scattering rate, similar to the
high-$T_{c}$ cuprates. Figures \ref{fig:CeCoIn5_fig1} (a) and (b)
depict $H$-dependence of $\kappa$ for {\boldmath $H$}$\parallel
ab$ ($H_{c2}\simeq 11$~T) and {\boldmath $H$}$ \perp ab$
($H_{c2}\simeq 5$~T) below $T_c$, respectively.   At all
temperatures,  $\kappa$ decreases with $H$, and the $H$-dependence
is less pronounced at lower $T$ in both configurations. For
{\boldmath $H$}$ \perp ab$, $\kappa$ has a discontinuous jump to
the normal state value at $H_{c2}$ below 1.0~K (see also the inset
of Fig.\ref{fig:CeCoIn5_fig1}(b)), indicating a first-order phase
transition. The data in Figs.\ref{fig:CeCoIn5_fig1} (a) and (b),
in which the $H$-dependence of $\kappa$ is more gradual with
decreasing $T$, are consistent with the picture where the
scattering of the field-induced QPs is the main origin for the
$H$-dependence of $\kappa$.

\begin{figure}[t]
\begin{center}
\includegraphics[width=8cm]{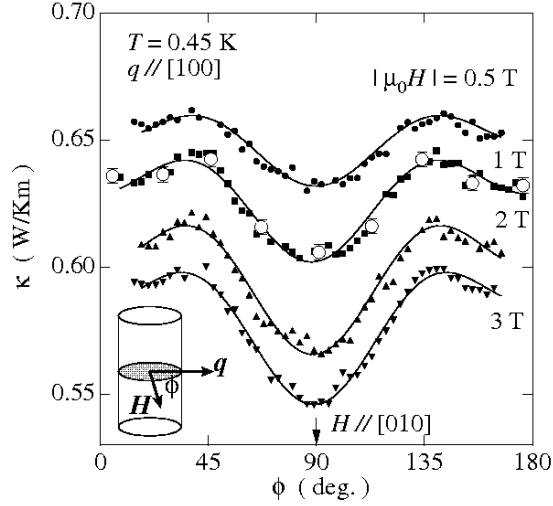}
\caption{The $a$-axis thermal conductivity $\kappa(H, \phi)$ ({\bf q}$\parallel a$) in {\boldmath $H$} rotated within the $ab$-plane as a function of $\phi$ for CeCoIn$_5$. $\phi$ is the angle between {\bf q} and {\boldmath $H$} (see the inset).   The solid lines represent the result of the fitting by the function $\kappa(H,\phi) = C_0 +C_{2\phi}\cos2\phi + C_{4\phi}\cos4\phi$, where $C_0$, $C_{2\phi}$ and $C_{4\phi}$ are constants.  The solid circles represent $\kappa(H, \phi)$ at $H$=1~T which are obtained under the field cooling condition at every angle. }
\label{fig:CeCoIn5_fig2}
\end{center}
\end{figure}

Given the complexity and richness of behavior of this
superconductor, it is natural to ask whether one can draw
conclusions about the symmetry of the gap from the analysis of the
thermal conductivity rooted essentially in a BCS-like theory. We
believe that the answer is affirmative. In the following it is
important to note that all the measurements are done far below the
upper critical field, i.e. far away from the possible competing
states, first order transition, and quantum critical behavior. The
thermal conductivity is measured at temperatures far below that of
the inelastic scattering-induced peak in $\kappa(T)$.
Consequently, we believe that, similar to the high-T$_c$
superconductors, the BCS-like model gives the semi-quantitatively
correct results in this regime.

Figure \ref{fig:CeCoIn5_fig2} displays $\kappa(H, \phi)$ as a
function of $\phi=$({\bf q}, {\boldmath $H$}) at
$T$=0.45~K of CeCoIn$_5$ \cite{izawaCe}. The solid circles in
Fig.\ref{fig:CeCoIn5_fig2} show $\kappa(H, \phi)$ at $H$=1~T which
are obtained under the field cooling at each angle.   In all data,
as shown by the solid lines in Fig.~2, $\kappa(H, \phi)$ can be
decomposed into three terms,
\begin{equation}
\kappa(\phi) = \kappa_{0} + \kappa_{2\phi} + \kappa_{4\phi},
\label{eqn:4fold}
\end{equation}
where $\kappa_{0}$ is a $\phi$-independent term, and
$\kappa_{2\phi} = C_{2\phi}\cos 2\phi$ and $\kappa_{4\phi} =
C_{4\phi}\cos 4\phi$ are terms with 2- and 4-fold symmetry with
respect to the in-plane rotation, respectively.

Figures \ref{fig:CeCoIn5_fig3} (a)-(d) display  $\kappa_{4\phi}$
(normalized by the normal state value $\kappa_n$).      It is
clear that $\kappa_{4\phi}$ exhibits a maximum at {\boldmath
$H$}$\parallel$[110] and [1$\overline{\rm 1}$0] at all
temperatures.    Figure 4  and the inset show the $T$- and $H$-
dependences of $|C_{4\phi}|/\kappa_n$. Below $T_c$ the amplitude
of $\kappa_{4\phi}$ increases gradually and shows a steep increase
below 1~K with decreasing $T$.  At low temperatures,
$|C_{4\phi}|/\kappa_n$ becomes greater than 2\%.

We note that the anisotropy of $H_{c2}$   ($H_{c2}^{\parallel
[100]}\simeq1.03H_{c2}^{\parallel [110]}$) is too small to explain
the large amplitude of $|C_{4\phi}|/\kappa_n>$ 2\% at $H \ll
H_{c2}$.   Further, and more importantly, the sign of the observed
fourfold symmetry is opposite to the one expected from the
anisotropy of $H_{c2}$.  The observed 4-fold symmetry above $T_c$
is extremely small; $|C_{4\phi}|/\kappa_n< $0.2~\%.  Thus the
anisotropies arising from $H_{c2}$ and the band structure are
incompatible with the data.  Note also that, even if there is a
fraction of electrons that remain uncondensed at low $T$, as was
recently suggested \cite{Tanatar05}, they can only influence the
twofold (via the orbital magnetoresistance), rather than the
fourfold symmetry. These considerations lead us to conclude that
{\it the 4-fold symmetry with large amplitude well below $T_c$
originates from the QP structure.}

\begin{figure}[t]
\begin{center}
\includegraphics[width=8cm]{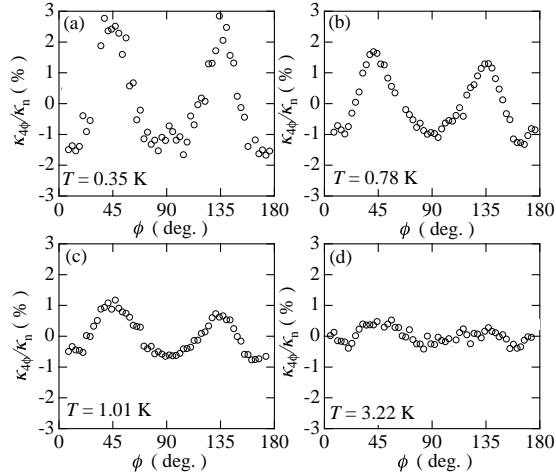}
\caption{(a)-(d) The 4-fold symmetry $\kappa_{4\phi}$ normalized by the normal state value $\kappa_n$ at several temperatures.  }
\label{fig:CeCoIn5_fig3}
\end{center}
\end{figure}

We now address the sign of the 4-fold symmetry.   For nodal lines
perpendicular to the layers, two effects compete in determining
$\kappa_{4\phi}$. The first is the DOS oscillation under  the
rotation of {\boldmath $H$} within the $ab$-plane. The second is
the quasiparticle scattering off the vortex lattice, which has the
same symmetry as the gap function
\cite{yu,aubin,vekhter2,Vorontsov1}. As discussed above, it is
likely that at $T\approx 0.25 T_c$ and $H\leq 0.25 H_{c2}$ the
second effect is dominant \cite{vekhter2,Vorontsov1}. In this
case, $\kappa$ attains the maximum value when {\boldmath $H$} is
along a node and has a minimum when {\boldmath $H$} is directed
towards the antinodal directions
\cite{maki1,Vorontsov1,yu,aubin,ocana}. The amplitude of the
four-fold symmetry $|C_{4\phi}|$ in quasi-2D $d$-wave
superconductors is roughly estimated to be a few percent of
$\kappa_n$ \cite{maki1,Vorontsov1}, which is of the same order as
the experimental results.

It is interesting to compare our results on CeCoIn$_5$  with the
corresponding results on YBa$_2$Cu$_3$O$_{7-\delta}$, in which the
4-fold symmetry has been reported in the regime where the Andreev
scattering dominates.  In YBa$_2$Cu$_3$O$_{7-\delta}$ with
$d_{x^2-y^2}$ symmetry, $\kappa_{4\phi}$ has maxima  at {\boldmath
$H$}$\parallel$[110] and [1${\rm \bar 1}$0] \cite{yu,aubin,ocana},
in accord with our CeCoIn$_5$ data.  Thus the sign of the present
fourfold symmetry indicates {\it the superconducting gap with
nodes located along the ($\pm\pi,\pm\pi$)-directions, similar to
the high-$T_c$ cuprates; CeCoIn$_5$  most likely belongs to the
$d_{x^2-y^2}$ symmetry. }

It is worth commenting on the gap symmetry of CeCoIn$_5$
determined from the other techniques.  Small angle neutron
scattering experiments have reported the square lattice of flux
lines \cite{eskild}, whose orientation relative to the crystal
lattice is consistent with the expectation for the
$d_{x^2-y^2}$-symmetry. Recent point contact spectroscopy, which
measured the Andreev reflection at the normal metal/CeCon$_5$
interface with two crystallographis orientations, (001) and (110),
have also concluded that the symmetry is $d_{x^2-y^2}$
\cite{WKPark2}. In contrast to these results, the angular
variation of the heat capacity  in {\boldmath H} rotated within
the $ab$-plane originally was interpreted as evidence for the
$d_{xy}$-symmetry \cite{aoki}. However, recent theoretical
analysis suggested that, when the redistribution of the spectral
density due to vortex scattering is accounted for, the specific
heat is also consistent with the $d_{x^2-y^2}$ gap
\cite{Vorontsov1}.

Therefore most of the measurements suggests that this material has
$d_{x^2-y^2}$ gap symmetry,  likely implying that the
antiferromagnetic fluctuations are important for
superconductivity. This observation qualitatively agrees with
recent NMR and neutron scattering experiments which reported
anisotropic spin fluctuations. While CeCoIn$_5$ is a very complex
system, we believe that the measurements of the field induced
anisotropy provide a strong evidence for the symmetry of the
superconducting gap.

\subsection{$\kappa$-(BEDT-TTF)$_2$Cu(NCS)$_2$}

The nature of the superconductivity in quasi-2D
$\kappa$-(BEDT-TTF)$_2$X salts [in the following abbreviated as
$\kappa$-(ET)$_2$X] , where the ion X can, for example, be
Cu(SCN)$_2$, Cu[N(CN)$_2$]Br or I$_3$, has attracted considerable
attention.   In these compounds, the large molecules are coupled,
forming narrow bands with low carrier density.  It is known that
ET molecules constitute a two dimensional conducting sheet in the
crystal $bc$-plane,  alternating with insulating layers of anions
X.   Within the conducting layer, ET molecules are arranged in
dimer pairs with alternating orientations (Fig.~26).   In
$\kappa$-(ET)$_2$Cu(NCS)$_2$, superconductivity occurs in
proximity to the AF ordered state in the phase diagram, implying
that the AF spin-fluctuations should play an important role for
the occurrence of superconductivity; some (but not all) of the
electronic properties of these superconductors are strikingly
similar to the high-$T_c$ cuprates \cite{mac,Schmalian}.
\begin{figure}[t]
\begin{center}
\includegraphics[width=8cm]{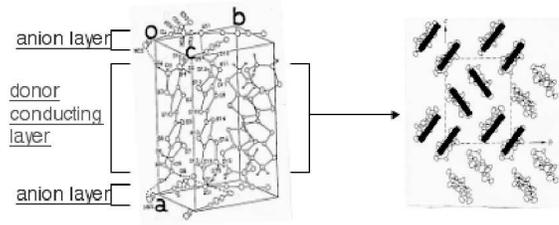}
\caption{Crystal structure of $\kappa$-(BEDT-TTF)$_2$Cu(NCS)$_2$}
\end{center}
\end{figure}

\begin{figure}[b]
\begin{center}
\includegraphics[width=7cm]{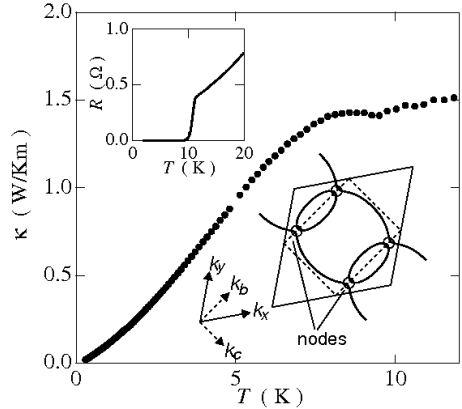}
\caption{Temperature dependence of the thermal conductivity in zero field.   The heat current {\bf q} was applied along the $b$-axis.   Upper inset: The resistive transition at $T_c$. Lower inset: The Fermi surface of
$\kappa$-(ET)$_2$Cu(NCS)$_2$.  The Fermi surface consists of quasi-1D and 2D hole pocket. The node directions determined in our experiment  are also shown. }
\label{fig:BEDT_fig1}
\end{center}
\end{figure}

The structure of the superconducting order  parameter of
$\kappa$-(ET)$_2$X salts has been examined by several techniques
\cite{kanoda}.  Results strongly favoring  $d$-wave pairing with
line nodes came from NMR \cite{maya,deso}, thermal conductivity,
and penetration depth \cite{carrington99,pinteric} experiments
\cite{belin} on X=Cu[N(CN)$_2$]Br and Cu(NCS)$_2$. The STM
\cite{nomura} and mm-wave transmission \cite{schrama} experiments
reported strong modulation of the gap structure, although they
arrived at very different conclusions regarding the nodal
directions.  In contrast to these experiments, specific heat
measurements on $\kappa$-(ET)$_2$Cu[N(CN)$_2$]Br near $T_c$
suggested a full gap\cite{elsinger}.

Figure \ref{fig:BEDT_fig1} depicts the  $T$-dependence of
$\kappa$. Since the phonon thermal conductivity, $\kappa^{ph}$,
dominates the electronic contribution, $\kappa^{el}$, near $T_c$,
the enhancement of $\kappa$ below $T_c$ reflects the increase of
the phonon mean free path as the electron pairs condense.
Figures~\ref{fig:BEDT_fig2}  (a) and (b) depict the $H$-dependence
of $\kappa$ in perpendicular ({\boldmath $H$}$\perp bc-$plane) and
parallel ({\boldmath $H$}$\parallel bc-$plane) field,
respectively. In the perpendicular field, $\kappa(H)$ shows a
monotonic decrease up to $H_{c2}$ above 1.6~K, which can be
attributed to the suppression of the phonon mean free path by the
introduction of the vortices \cite{belin,VekhterC}.  Below 1.6~K,
$\kappa(H)$ exhibits a dip below $H_{c2}$.  The minimum of
$\kappa(H)$ arises from the competition between $\kappa^{ph}$,
which always decreases with $H$, and $\kappa^{el}$, which
increases steeply near $H_{c2}$.  Consequently the magnitude of
the increase of $\kappa(H)$ below $H_{c2}$ provides a lower limit
of the electronic contribution, which grows rapidly below 0.7~K;
$\kappa^{el}_n/\kappa_n$ is roughly estimated to be $\gtrsim$5\%
at 0.7~K and $\gtrsim$15\% at 0.42~K, where $\kappa^{el}_n$ and
$\kappa_n$ are the electronic and total thermal conductivity in
the normal state above $H_{c2}$, respectively.

\begin{figure}[t]
\begin{center}
\includegraphics[width=7cm]{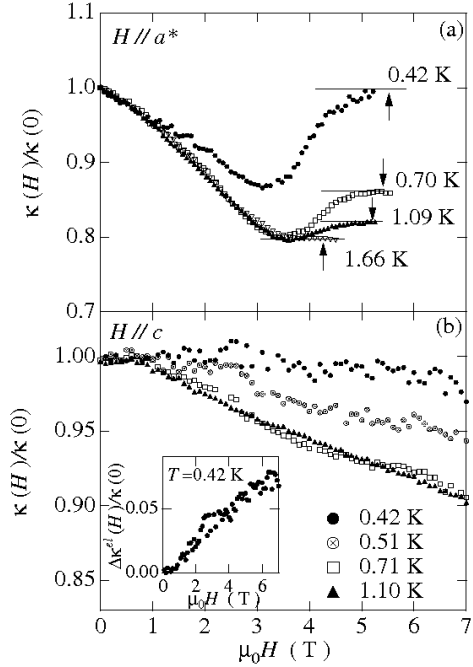}
\caption{Field dependence of the in-plane thermal conductivity (a) in perpendicular
and (b) in parallel field ({\boldmath $H$} $\parallel c$) at low temperatures.
Deviation from the horizontal line shown by arrows marks $H_{c2}$ .
Inset:  Field dependence of the electronic  thermal conductivity $\Delta \kappa^{el}$ in parallel field. }
\label{fig:BEDT_fig2}
\end{center}
\end{figure}

We now move on  to the angular variation of $\kappa$ as {\boldmath
$H$} is rotated within the 2D $bc$-plane.  Figures
\ref{fig:BEDT_fig3} (a)-(c) display $\kappa$({\boldmath $H$},
$\phi)$ as a function of $\phi=$({\bf q}, {\boldmath $H$}) at low
temperatures.   Above 0.72~K $\kappa(${\boldmath $H$}, $\phi)$
shows a minimum at $\phi=90^{\circ}$, indicating simply that the
transport is better for the heat current parallel to the vortices.
On the other hand, at lower temperatures, the angular variation
changes dramatically, exhibiting a double minimum as shown in
Figs.\ref{fig:BEDT_fig3} (b) and (c). In all data, we fit $\kappa
(\phi)$ as a sum of three terms: a constant, a two fold,
$\kappa_{2\phi}$, and a fourfold, $\kappa_{4\phi}$. These fits are
shown by the solid lines in Figs.\ref{fig:BEDT_fig3} (a)-(c).
Since a large twofold symmetry is observed even above 0.7~K, where
$\kappa^{ph}$ dominates, $\kappa_{2\phi}$ is mainly phononic in
origin.  In what follows, we will address the fourfold symmetry
which is directly related to the electronic properties.

Figures \ref{fig:BEDT_fig3}  (d)-(f) display $\kappa_{4\phi}$
normalized by $\kappa_n$ \cite{izawaET}.  At $T$=0.72~K, the
fourfold component is small: $|C_{4\phi}|/\kappa_n<0.1$\%.  On the
other hand, a clear fourfold component with
$|C_{4\phi}|/\kappa_n\sim 0.2$\% is resolved at 0.52 and 0.43~K.
As discussed before, the contribution of $\kappa^{el}$ grows
rapidly below 0.7~K and contitutes a substantial portion of the
total $\kappa$ at 0.4~K. Therefore it is natural to consider that
{\it the fourfold oscillation is purely electronic in origin.}
Although $|C_{4\phi}|$ at 0.42~K is as small as 0.2\% in
$\kappa_n$, it is likely close to 1.5-2\% of $\kappa^{el}_n$ and
is also a few per cent of $\kappa^{el}(0)$, assuming
$\kappa^{el}_n/\kappa_n\sim 0.15$. The band structure of the
crystal is very unlikely to be an origin of the fourfold symmetry,
as the Fermi surface of this material is nearly elliptic with
twofold symmetry, and the fourfold modulation is negligible, if
present at all \cite{ohshima}.  Hence we conclude that {\it the
observed fourfold symmetry originates from the superconducting gap
nodes.}

\begin{figure}[t]
\begin{center}
\includegraphics[width=9cm]{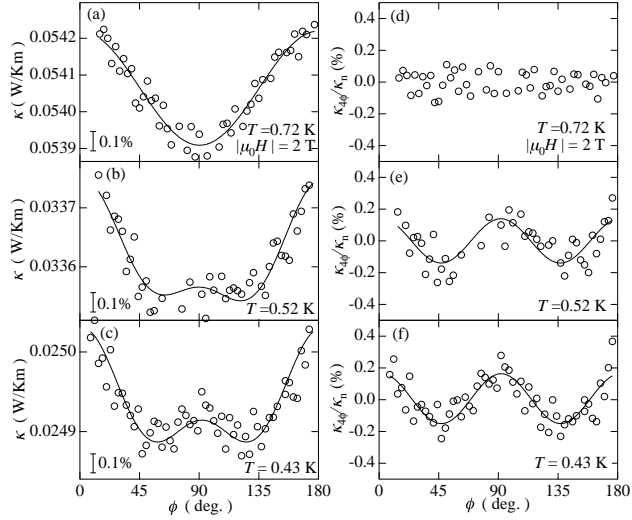}
\caption{(a)-(c)Angular variation of
$\kappa(${\boldmath $H$},$\phi)$ in $|\mu_{0}${\boldmath $H$}$|$=2~T for different temperatures.
$\phi$ is the angle between {\bf q} and {\boldmath $H$}.
 The solid lines represent the result of the fitting by the function
 $\kappa(H,\phi) = C_0 +C_{2\phi}\cos2\phi + C_{4\phi}\cos4\phi$,
 where $C_0$, $C_{2\phi}$ and $C_{4\phi}$ are constants.
 (d)-(f) The fourfold symmetry $\kappa_{4\phi}$ obtained from  (a)-(c).  }
\label{fig:BEDT_fig3}
\end{center}
\end{figure}
The main question is whether the anisotropy of  $\kappa^{el}$ is
associated with the DOS oscillation or Andreev scattering off the
vortices. From general arguments \cite{vekhter2,Vorontsov1}, the
density of state effects dominate at low $T$, such as our
experimental temperature of $T_c$/30. Also, since $\kappa^{el}$
increases with $H$, as shown in the inset of Fig.~2(b), we believe
that the DOS enhancement underlies the $H$-dependence of
$\kappa^{el}$ at 0.42~K.

If the DOS oscillations indeed dominate, $\kappa^{el}$ attains its
maximum value when {\boldmath $H$} is directed along the antinodal
directions, and has a minimum when {\boldmath $H$} is along the
nodes. According to Refs.\cite{Vorontsov1,won}, $|C_{4\phi}|$ in
the $d$-wave superconductors arising from the DOS oscillation is
roughly estimated to be a few percent of $\kappa^{el}(0)$, which
is in the same order to the experimental results. Since
$\kappa_{4\phi}$ exhibits a maximum when {\boldmath $H$} is
applied parallel to the $b$- and $c$-axes of the crystal,
we conclude that {\it the gap nodes are along the directions
rotated 45$^{\circ}$ relative to the $b$- and $c$-axes};
the nodes are situated near the band gap between the 1D and 2D
bands (see the upper inset of Fig.~1).  This result is consistent
with the STM experiments \cite{nomura}.

We emphasize that {\it the determined nodal  structure is
inconsistent with the recent theories based on the AF spin
fluctuations}.  In the AF spin fluctuation scenario, it is natural
to expect the nodes to be along the $b$- and $c$- directions
since the AF ordering vector is along the {\bf b}-axis, which
provides partial nesting.  If we take the same conventions for the
magnetic Brillouin zone as in the high-$T_c$ cuprates with
$d_{x^2-y^2}$ symmetry (see Fig.~1 (c) in Ref.\cite{Schmalian}),
the superconducting gap symmetry of $\kappa$-(ET)$_2$Cu(NCS)$_2$
is $d_{xy}$.  Recently, it has been suggested that the nodal
structure depends on the hopping integral between ET molecules,
even if the superconductivity is mediated by AF fluctuation.  For
instance, $d_{xy}$ symmetry dominates over $d_{x^2-y^2}$ when the
dimerization of ET molecules is not too strong, which appears to
be the case for $\kappa$-(ET)$_2$Cu(NCS)$_2$ \cite{kuroki}.
Moreover, when the second nearest neighbor hopping integral $t_b$
between the dimer is comparable to the nearest  neighbor hopping
integral $t_c$ $(t_b \sim t_c)$, the $d_{xy}$-symmetry is
stabilized \cite{moriya}. Indeed, in $\kappa$-(ET)$_2$Cu(NCS)$_2$,
$t_b/t_c$ is 0.8, which is close to unity.  It has also been
argued that the charge fluctuations rather than spin fluctuations
may be relevant to the unconventional superconductivity in
$\kappa$-(ET)$_2$Cu(NCS)$_2$. Consequently, our results for the
gap symmetry should serve as a constraint on future development of
the theories of superconductivity in this family of compounds.

\begin{figure}[b]
\begin{center}
\includegraphics[width=5cm]{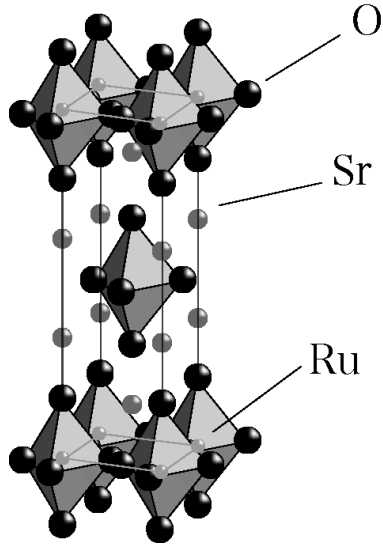}
\caption{Crystal structure of Sr$_2$RuO$_4$.}
\end{center}
\end{figure}

\subsection{Sr$_2$RuO$_4$}

Ever since its discovery in 1994 \cite{maeno1994},  the
superconducting properties of the layered ruthenate Sr$_2$RuO$_4$
attracted considerable interest \cite{rice}.  The crystal
structure of this material is same as La$_2$CuO$_4$, a parent
compound of high-$T_c$ cuprates (Fig.~30).

The superconducting state of Sr$_2$RuO$_4$ stimulated great
interest because NMR Knight shift remains unchanged from the
normal state value below $T_c$, indicating that the pairing state
may be a spin triplet \cite{ishidaSr}. Recent phase sensitive
experiments are controversial: although odd-parity of the
superconducting wave function was suggested by Ref.\cite{nelson},
it was also claimed that the result of Ref.\cite{nelson} can be
interpreted in the even parity framework \cite{mazin}.    $\mu$SR
measurements report the appearance of  static spontaneous magnetic
field below $T_c$, which can be interpreted as a sign of broken
time reversal symmetry \cite{mSRSRO}. The specific heat
$C_p$\cite{nishizaki}, NMR relaxation rate \cite{ishida2}and
thermal conductivity \cite{suzuki} indicate the presence of nodal
lines in the superconducting gap.  These results have motivated
theorists to propose new models for the superconductivity in the
ruthenates \cite{miyake,hasegawa,graf,ogata}. We, of course,
address this issue here from the standpoint of the thermal
conductivity measurements.

 \begin{figure}[t]
\begin{center}
\includegraphics[width=7cm]{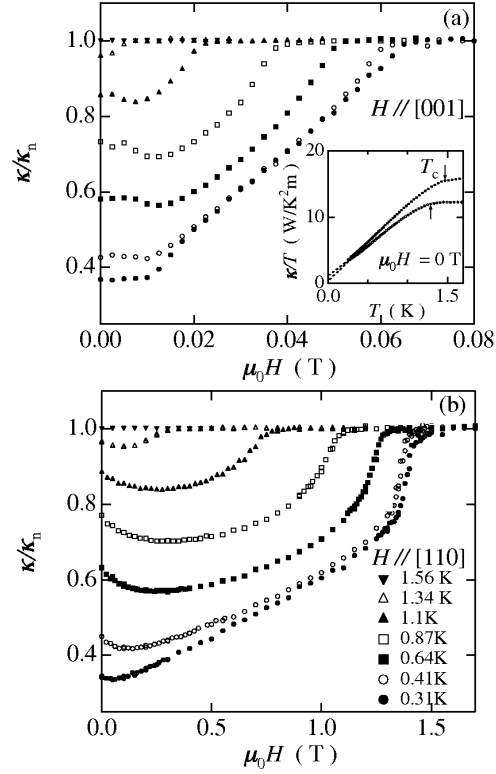}
\caption{Field dependence of the in-plane thermal conductivity ({\bf q}$\parallel$[110]) of Sr$_2$RuO$_4$ ($T_c$=1.45~K) in (a)perpendicular field {\boldmath $H$}$\perp ab$-plane and  (b)parallel field {\boldmath $H$}$\parallel$[110].  In perpendicular field, $\kappa$ is indepedent of $H$ below the lower critical fields.  Inset:  Temperature dependence of $\kappa/T$ in zero field for two crystals with different $T_c$ ($T_c$=1.45 and 1.32~K).}
\label{fig:Sr2RuO4_fig1}
\end{center}
\end{figure}

 \begin{figure}[t]
\begin{center}
\includegraphics[width=9cm]{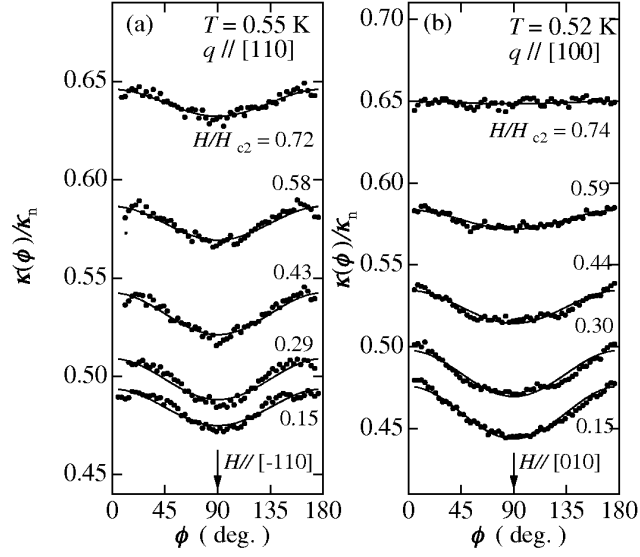}
\caption{(a) In-plane thermal conductivity normalized by the normal state value in {\boldmath $H$} rotated within the $ab$-plane as a function of $\phi$ of Sr$_2$RuO$_4$ ($T_c$=1.45~K).  {\bf q} is applied to the [110]-direction.  (b) The same plot for the sample with $T_c$=1.37~K.   {\bf q} is applied to the [100]-direction.  $\phi$ is the angle between {\boldmath $H$} and  {\bf q}.  The solid lines represent the twofold component in $\kappa(\phi)/\kappa_n$.  }
\label{fig:Sr2RuO4_fig2}
\end{center}
\end{figure}

 \begin{figure}[b]
\begin{center}
\includegraphics[width=9cm]{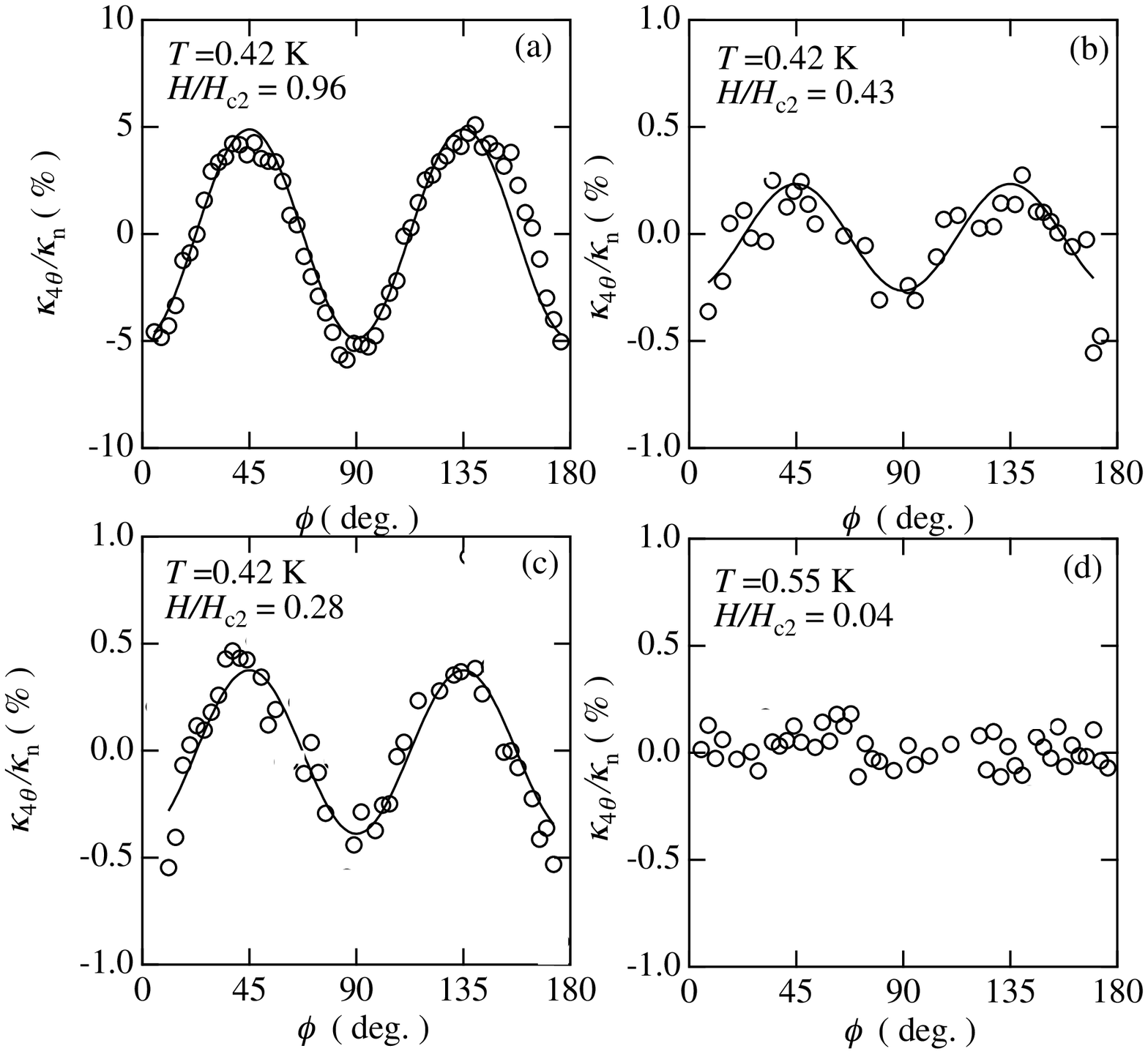}
\caption{(a)-(d) The fourfold symmetry $\kappa_{4\phi}/\kappa_n$ at several fields obtained from Fig.~\ref{fig:Sr2RuO4_fig2} (a).  }
\label{fig:Sr2RuO4_fig3}
\end{center}
\end{figure}

Inset of Fig.\ref{fig:Sr2RuO4_fig1}(a)  shows the $T$-dependence
of $\kappa/T$ in zero field for crystals with $T_c$=1.35~K and
1.5~K.  The system is very pure, as is clear from the electrical
resistivity of the order of 0.1$\mu\Omega\cdot$cm. At low $T$ the
electronic contribution to the thermal conductivity is dominant.
At the superconducting transition, $\kappa/T (H)$ shows a kink.
Figures \ref{fig:Sr2RuO4_fig1} (a) and (b) show the $H$-dependence
of $\kappa$ for the sample with $T_c$=1.45~K in perpendicular
({\bf H}$\perp ab$-plane) and parallel fields ({\bf H}$\parallel
ab$-plane), respectively.  In both orientations, $\kappa$
increases with $H$ after an initial decrease at low fields.  The
subsequent minimum is much less pronounced at lower temperatures.
At low $T$, $\kappa$ increases linearly with $H$. For the in-plane
field $\kappa$ rises very rapidly as $H$ approaches $H_{c2}$ and
attains its normal value with a large slope ($d\kappa/dH$), while
$\kappa$ in perpendicular field remains linear in $H$ up to
$H_{c2}$. In superconductors the slope of  $\kappa(H)$ below
$H_{c2}$ increases with purity of the sample \cite{VekhterC}, so
that the data for the in-plane field suggest a clean limit. A
rough estimate can be done as follows:  let $\Gamma$ be the pair
breaking parameter estimated from the Abrikosov-Gorkov equation
$\Psi(1/2+\Gamma/2 \pi T_c)-\Psi(1/2)=\ln(T_{c0}/T_c)$, where
$\Psi$ is a digamma function and $T_{c0}$ is the transition
temperature in the absence of the pair breaking.  Assuming
$T_{c0}$=1.50~K and $\Delta=1.76T_c$, $\Gamma/\Delta$ is estimated
to be 0.025 (0.067) for $T_c$=1.45~K($T_c$=1.37~K). Thus the
dependence of $\kappa({\boldmath H)}$ observed in very clean
crystals is consistent with the $\kappa$ of superconductors with
line nodes.

We now discuss the angular variation of the  thermal conductivity.
Figures~\ref{fig:Sr2RuO4_fig2} (a) and (b) depict
$\kappa(${\boldmath $H$}, $\phi)$ as a function of $\phi=$ ({\boldmath
$q,H$}) \cite{izawaSr}.   In all the data $\kappa({\boldmath
H},\phi)$ can decomposed as in Eq.(15).  Figures
\ref{fig:Sr2RuO4_fig3} (a)-(d) show $\kappa_{4\phi}/\kappa_n$ as a
function of $\phi$ after the subtraction of $\kappa_{0}$- and
$\kappa_{2\phi}$-term from $\kappa$.    Figure
\ref{fig:Sr2RuO4_fig4} depicts the $H$-dependence of
$|C_{4\phi}|$.  In the vicinity of $H_{c2}$ where $\kappa$
increases steeply, $|C_{4\phi}|/\kappa_n $ is of the order of a
several percent (see Fig.~\ref{fig:Sr2RuO4_fig3}(a)). We point out
that both the sign and amplitude of  $C_{4\phi}$ in the vicinity
of $H_{c2}$ is mainly due to the in-plane anisotropy of $H_{c2}$.
In Fig.\ref{fig:Sr2RuO4_fig4}, we plot the amplitude of the
fourfold oscillation calculated from the in-plane anisotropy of
$H_{c2}$. The calculation reproduces the data.

 \begin{figure}[t]
\begin{center}
\includegraphics[width=7cm]{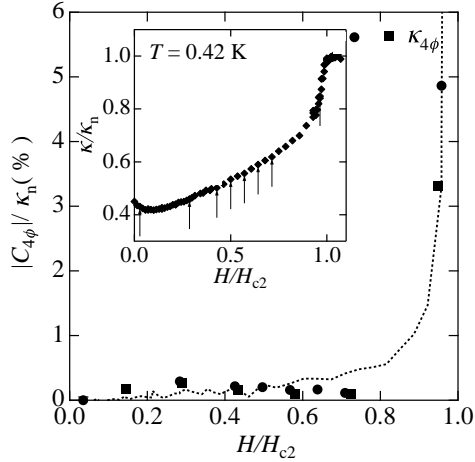}
\caption{The amplitude of four fold symmetry as a function of
$H/H_{c2}$.  The filled circles and squares indicate
$|C_{4\theta}|/\kappa_{n}$ at $T$=0.42~K and 0.55~K, respectively.
  The solid line represents
$|C_{4\theta}|/\kappa_n$ calculated from the 4-fold symmetry of
$H_{c2}$.  Inset: $H$-dependence of $\kappa$.  The arrows indicates
the points we measured  $C_{4\theta}$. }
\label{fig:Sr2RuO4_fig4}
\end{center}
\end{figure}

$|C_{4\phi}|/\kappa_n $ decreases rapidly  and is about 0.2-0.3\%
at lower field where $\kappa$ increases linearly with $H$ (see
Figs.~\ref{fig:Sr2RuO4_fig3} (b) and (c)).  At very low field
where $\kappa$ decreases with $H$, no discernible 4-fold
oscillation is observed within the resolution of
$|C_{4\phi}|/\kappa_n<$ 0.1\% (see
Fig.~\ref{fig:Sr2RuO4_fig3}(d)).  Thus the amplitude of the
fourfold oscillation at low field of Sr$_2$RuO$_4$ is less than
1/20 of those in CeCoIn$_5$ and $\kappa$-(ET)$_2$Cu(NCS)$_2$.
These results lead us conclude that the line nodes are not located
perpendicular to the plane, but located parallel ot the plane,
{\it i.e.} horizontal node.  If we accept the spin triplet
superconducticity with broken time reversal symmetry,  the gap
symmetry which is most consistent with the in-plane variation of
thermal conductivity is $\mbox{\bf d}(\mbox{\bf k}) = \Delta_{0}\mbox{\boldmath
$\widehat{z}$}(k_{x}+ik_{y})(\cos ck_z+\alpha)$, in which the
substantial portion of the Cooper pairs occurs between the
neighboring RuO$_2$ planes.  Similar conclusion was obtained from
the $c$-axis thermal conductivity measurements \cite{tanatar}.
Recently, the in plane-variation of heat capacity has been
reported down to 100~mK.  Below 200 mK, small but finite four-fold
oscillation, which disappears at low $H$, was observed
\cite{deguchi1,deguchi2}.   This indicates that the gap function
has a modulation around the $c$-axis, though finite gap remains.

At an early stage, the gap symmetry of  Sr$_2$RuO$_4$ was
discussed in analogy with $^3$He, where the Cooper pairs are
formed by the exchange of ferromagnetic spin fluctuations
\cite{rice}.   However the inelastic neutron scattering
experiments showed the existence of strong incommensurate
antiferromagnetic correlations,  and no sizable ferromagnetic spin
fluctuations, indicating that the origin of the triplet pairing is
not simply a ferromagnetic interaction \cite{sidis}.  The present
results impose strong constraints on models that attempt to
explain the mechanism of the triplet superconductivity.  We
finally comment on the orbital-dependent superconductivity
scenario, in which three different bands have different
superconducting gaps \cite{agta}.  In this case, our main
conclusion can be applicable to the band with the largest gap
(presumably the $\gamma$-band).

\section{Summary}

In this paper,  we discussed ''how to determine the
superconducting gap structure in the bulk?". We show that the
measurements thermal conductivity and specific heat in magnetic
field rotating various directions relative to the crystal axis can
determine the position and type of nodes. In
Fig.~\ref{fig:gapfunctions}, we summarize the nodal structure of
several unconventional superconductors, including  borocarbide
YNi$_2$B$_2$C \cite{izawaYN2}, heavy fermions
UPd$_2$Al$_3$\cite{watanabe}, CeCoIn$_5$\cite{izawaCe}, and
PrOs$_4$Sb$_{12}$\cite{izawaPr}, organic superconductor,
$\kappa$-(BEDT-TTF)$_2$Cu(NCS)$_2$\cite{izawaET}, and ruthenate
Sr$_2$RuO$_4$ \cite{izawaSr}, which are determined by the present
technique. While more theoretical work is clearly needed to arrive
at a more quantitative description of the data, we feel confident
that the method provides a uniquely powerful route towards
determination of the nodal structure in novel superconductors.

 \begin{figure*}[t]
\begin{center}
\includegraphics[width=11cm]{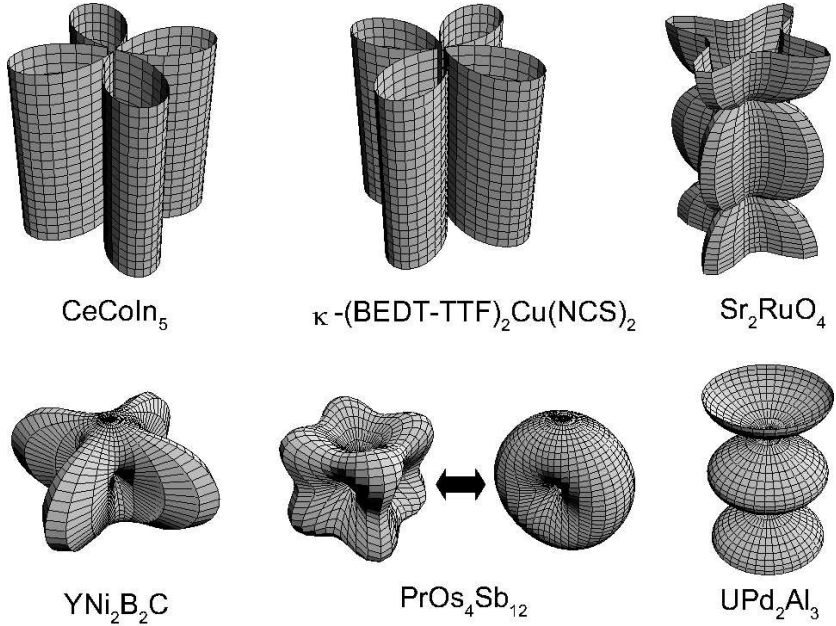}
\caption{ The nodal structure of several unconventional superconductors,
including quasi-two dimensional heavy fermion CeCoIn$_5$, organic $\kappa$-(BEDT-TTF)$_2$Cu(NCS)$_2$,  Ruthenate Sr$_2$RuO$_4$(The gap
structure takes into account additional fourfold
modulation as indicated by the specific heat measurements.), borocarbide YNi$_2$B$_2$C, heavy fermion PrOs$_4$Sb$_{12}$, and heavy fermion UPd$_2$Al$_3$, which are determined by angular variation of the thermal conductivity }
\label{fig:gapfunctions}
\end{center}
\end{figure*}

\ack
This work has been done in collaboration with  J.~Goryo,
K.~Kamata, Y.~Kasahara, Y.~Nakajima, T.~Watanabe (Univ. of Tokyo), P.~Thalmeier (Max-Planck Institute), and K.~Maki (Univ. of Southern
California).  Samples were provided by Y.~Onuki, R.~Settai, H.~Shishido,Y.~Yoshida
(Osaka Univ.), Y.~Haga (Japan Atomic Research Institute),
T.~Sasaki (Tohoku Univ.), M.~Nohara, H.~Takagi (Univ. of
Tokyo), S.~Osaki, H.~Sugawara, H.~Sato (Tokyo Metropolitan Univ.).
    We thank K.~Behnia, H.~Harima, P. J.~Hirschfeld, M.~Ichioka, K.~Ishida,
H.~Kontani, K.~Kuroki, H.~Kusunose, J.P.~Pascal, T.~Sakakibara,  T.~Shibauchi,  K.~Machida, V.P.~Mineev,
P.~Miranovic, Y.~Maeno,   N.~Nakai, M.~Ogata,  Y.~Tanaka, M.~Tanatar,
M.~Udagawa, A.~Vorontsov, H.~Won, K.~Yamada and Y.~Yanase for valuable
discussions. This work was partly supported by Grants-in-Aid for
Scientific Reserch from MEXT (Y. M.) and the Board of
Regents of Louisiana (I. V.).


\end{document}